\documentclass[12pt, a4paper]{article}
\usepackage[margin=1in]{geometry}
\usepackage{amsfonts, amsmath, amssymb, bm}
\usepackage{tabu}
\usepackage{mathtools}
\usepackage[utf8x]{inputenc}
\usepackage{enumerate}
\usepackage{tgcursor}
\usepackage{xcolor}
\usepackage{graphicx}
\usepackage{placeins}
\usepackage{listings}
\usepackage{stackengine}
\usepackage{caption}
\usepackage{subcaption}
\usepackage{mathptmx}
\usepackage{setspace}
\usepackage{footnote}
\usepackage{comment}
\usepackage{footmisc}
\usepackage{hyperref}
\numberwithin{equation}{section}
\usepackage{eucal}
\usepackage{verbatim}
\usepackage{mathrsfs}
\usepackage{natbib}

\numberwithin{equation}{section}
\newtheorem{theorem}{Theorem}
\newtheorem{proposition}[theorem]{Proposition}

\DeclareMathAlphabet\EuScript{U}{eus}{m}{n}
\SetMathAlphabet\EuScript{bold}{U}{eus}{b}{n}

\renewcommand{\vec}[1]{\boldsymbol{#1}}
\newcommand*{\bdot}[1]{\overset{\mbox{\Large .}}{#1}}

\newcommand{\va}{\mathbf{a}}
\newcommand{\vA}{\mathbf{A}}
\newcommand{\vb}{\mathbf{b}}
\newcommand{\vB}{\mathbf{B}}
\newcommand{\vc}{\mathbf{c}}
\newcommand{\vC}{\mathbf{C}}

\newcommand{\vD}{\mathbf{D}}

\newcommand{\vf}{\mathbf{f}}
\newcommand{\vF}{\mathbf{F}}
\newcommand{\vg}{\mathbf{g}}
\newcommand{\vG}{\mathbf{G}}

\newcommand{\vH}{\mathbf{H}}

\newcommand{\vI}{\mathbf{I}}

\newcommand{\vM}{\mathbf{M}}

\newcommand{\vs}{\mathbf{s}}

\newcommand{\vu}{\mathbf{u}}
\newcommand{\vU}{\mathbf{U}}
\newcommand{\vv}{\mathbf{v}}

\newcommand{\vX}{\mathbf{X}}
\newcommand{\vy}{\mathbf{y}}

\newcommand{\vY}{\mathbf{Y}}

\newcommand{\vZ}{\mathbf{Z}}

\newcommand{\vepsilon}{\boldsymbol{\epsilon}}

\newcommand{\veta}{\boldsymbol{\eta}}

\newcommand{\vgamma}{\boldsymbol{\gamma}}

\newcommand{\vomega}{\boldsymbol{\omega}}

\newcommand{\vphi}{\boldsymbol{\phi}}
\newcommand{\vPhi}{\boldsymbol{\Phi}}
\newcommand{\vpi}{\boldsymbol{\pi}}

\newcommand{\vpsi}{\boldsymbol{\psi}}
\newcommand{\vPsi}{\boldsymbol{\Psi}}

\newcommand{\vsigma}{\boldsymbol{\sigma}}
\newcommand{\vSigma}{\boldsymbol{\Sigma}}

\newcommand{\vtheta}{\boldsymbol{\theta}}
\newcommand{\vTheta}{\boldsymbol{\Theta}}

\hypersetup{
    colorlinks=true,
    linkcolor=blue,
    urlcolor=blue,
    citecolor=black
}

\begin{document}

\begin{center}
   \Large\centering\normalfont{A Bayesian Approach for Spatio-Temporal Data-Driven Dynamic Equation Discovery}
\end{center}

\begin{center}
    Joshua S. North\footnote[1]{Corresponding author: jsnorth@lbl.gov, Earth and Environmental Sciences, Lawrence Berkeley National Laboratory, Berkeley, CA, 1 Cyclotron Road}, 
    Christopher K. Wikle\footnote[2]{wiklec@missouri.edu, Department of Statistics, University of Missouri, Columbia, MO, 146 Middlebush Hall}, 
    Erin M. Schliep\footnote[3]{emschlie@ncsu.edu, Department of Statistics, North Carolina State University, Raleigh, NC, 2311 Stinson Drive}
\end{center}

\begin{abstract}
Differential equations based on physical principals are used to represent complex dynamic systems in all fields of science and engineering.
Through repeated use in both academics and industry, these equations have been shown to represent real-world dynamics well.
Since the true dynamics of these complex systems are generally unknown, learning the governing equations can improve our understanding of the mechanisms driving the systems.
Here, we develop a Bayesian approach to data-driven discovery of non-linear spatio-temporal dynamic equations.
Our approach can accommodate measurement noise and missing data, both of which are common in real-world data, and accounts for parameter uncertainty.
The proposed framework is illustrated using three simulated systems with varying amounts of observational uncertainty and missing data and applied to a real-world system to infer the temporal evolution of the vorticity of the streamfunction.
    \begin{center}
    \end{center}
    
\end{abstract}

\section{Introduction}\label{sec:introduction}


Dynamic equations (DE) parameterized by partial differential equations (PDE) -- an equation relating a partial derivative of a variable to a function of its current state -- are used across all fields of science and engineering to describe complex processes.
DEs encode physical processes by a set of mathematical equations, enabling complex systems such as the spread of infectious disease \citep{Bolker1995, Mangal2008, Kuhnert2014}, evolution of invasive species \citep{Hastings1996, Liu2019}, weather and climate \citep{Charney1950, Holton2012}, and the flow of fluids \citep{White2006} to be characterized and modeled \citep[see also][for further discussion]{Higham2016a}.
Given that any real world process is only approximately characterized by mathematical relationships, mathematically derived DEs are inherently unable to completely characterize a real world system.
This suggests the need to use observations of the real world system to better characterize the underlying DEs.

Recently, there has been a push to use data to discover the governing equations in these complex systems.
Originally proposed using symbolic regression \citep{Bongard2007, Schmidt2009}, the focus has since shifted to either sparse regression or deep modeling.
The original sparse regression approach, termed \textit{Sparse Identification of Nonlinear Dynamics} \citep[SINDy;][]{Brunton2016}, used numerical differentiation to construct a response that is regressed against a library of functions that potentially govern the system.
Through sparse regression, either with an $\ell_1$ penalization term \citep{Tibshirani1996} or using a thresholding approach \citep{Zheng2019, Champion2020}, key terms governing a variety of ordinary differential equations (ODE) are identified.
Using the fundamental idea of SINDy, the framework was extended to include PDEs and parametric forms \citep{Schaeffer2017, Rudy2017, Rudy2019, Rudy2019a}, stochastic dynamical systems \citep{Boninsegna2018}, uncertainty quantification of the parameters \citep{Zhang2018b, Yang2019, Niven2020, Fasel2021, Hirsh2021}, and has been incorporated into a Python package \citep[PySINDy;][]{DeSilva2020}.

Deep models used for data-driven discovery of dynamics can broadly be grouped into two categories -- approximating dynamics \citep{Raissi2017, Raissi2018, Raissi2020, Sun2019, Wu2020} and discovering dynamics \citep{Both2021, Xu2019, Xu2020, Xu2021, Long2017, Long2019}.
Using deep models to approximate the dynamics of complex systems enables a computationally inexpensive method to obtain measurements of otherwise difficult to simulate systems while still obeying physical principles \citep[see][for an in-depth discussion on the topic]{Reichstein2019}.
However, our goal is the discovery of the governing equations where ``data-driven discovery'' refers to the discovery of the \textit{functional form} of the system and deep models have also been used in this context.
Combining deep modeling and sparse identification, \citet{Both2021} approximate the PDE using a neural network, which is used to compute derivatives and construct a sparse formulation similar to the SINDy approach.
\citet{Long2017, Long2019} use a symbolic neural network, an extension of symbolic regression, and a numerical approximation of differential operators in a feed-forward network to discover PDEs and \citet{Xu2021} use a fully connected neural network with a genetic algorithm to express and generate terms of a PDE.

Two open problems in data-driven discovery are (i) accounting for measurement uncertainty (i.e., missing data and measurement noise) and (ii) parameter uncertainty.
Existing methods that extend the SINDy framework to account for uncertainty quantification employ either a bootstrap approach \citep{Fasel2021} or a Bayesian approach with variable shrinkage/selection priors placed on the coefficients associated with the library terms \citep{Zhang2018b, Niven2020, Hirsh2021}.
These approaches directly follow the first steps of the SINDy framework, where derivatives are computed numerically, data is de-noised, and the feature library is constructed.
In this manner, the true uncertainty associated with the observed data is ignored; the estimate for the system uncertainty is now dependent on the numerical differentiation method, which subsequently influences the estimate of the parameter uncertainty.
\citet{Yang2020} developed a method to jointly account for uncertainty in the observed data and parameters based on differential Bayesian programming.
While this approach directly accounts for measurement uncertainty, it requires derivatives be computed using a numerical solver (e.g., Runge-Kutta), which can lead to numerical instabilities and cannot accommodate missing data.

To account for observational uncertainty and missing data when modeling complex non-linear systems, statisticians have incorporated dynamic equations parameterized by PDEs into Bayesian hierarchical models \citep[BHM;][]{Berliner1996, Royle1999, Wikle2001}.
These models, sometimes called physical statistical models (PSM), enable modeling mechanistic relationships within a probabilistic framework \citep[see][for an overview]{Berliner2003, Cres, Kuhnert2017}.
PSMs are composed of three sub-models -- data, process, and parameter models.
To account for observational and mechanistic uncertainty, PSMs consider the dynamics to be latent in the process stage, and represent the observed data in the data stage conditioned on these latent dynamics.
While PSMs have been used to model and better understand complex systems, such as ocean surface winds \citep{Wikle2001, Milliff2011} and the spread of avian species \citep{Wikle2003a, Hooten2008}, they require the dynamic relationships (although not weights/parameters associated with those relationships) to be specified \textit{a priori}.
To increase flexibility for representing complex processes, PSMs consider the  parameters that describe the influence of dynamic components to be random, and often allow them to have spatial or temporal dependence, enabling the model to adapt to the data.
While PSMs are adaptable to a variety of problems and provide inference on how the process may be evolving, they cannot be used to discover new dynamical relationships.

Recently, \citet{North2022} proposed a Bayesian data-driven discovery method that accounts for observational and parameter uncertainty using a BHM framework composed of data, process, and parameter models for ODEs.
Analogous to PSMs, the dynamics are modeled as a latent process and observational error is accounted for in the data model.
Allowing the dynamics to be a latent random process is different than previous data-driven discovery methods that attempt to quantify uncertainty. 
To link the dynamic system to its derivatives probabilistically, the dynamic process and all the derivatives are modeled using a basis expansion with a common set of basis functions.
Derivatives are then obtained analytically using the basis expansion, which incorporates dependence between the dynamic process and its derivatives.
A library of potential functions can be constructed based on the basis coefficients and functions, and a variable selection prior is used to identify the key functions governing the non-linear system.

Here, we propose a spatio-temporal extension to Bayesian data-driven discovery for PDEs.
The general framework follows \citet{North2022}, however the addition of the spatial dimension requires a reformulation of the process model.
To account for the extra dimension (i.e., space), we model the dynamic process as a higher-order tensor where the dimensions represent space, time, and the number of components (sometimes called the system states) in the system.
The tensor is decomposed using differentiable basis functions in space and time, probabilistically linking the dynamic system with its spatial and temporal derivatives.
The basis decomposition is incorporated into the BHM, enabling potential functions to be constructed using the basis functions and coefficients.
A variable selection prior on the coefficients produces a sparse solution set and the resulting system.
In contrast to the ODE discovery problem, the library of potential functions for PDEs can exhibit strong multicollinearity.
An additional contribution of this work is an approach to account for multicollinearity in high dimensional, dynamic systems.

We demonstrate our method using data generated from Burgers' equation, the heat equation, and a predator-prey reaction-diffusion equation with varying levels of measurement noise.
In addition, we demonstrate our model's ability to accommodate missing data using Burgers' equation.
The simulations show that our approach is robust to measurement noise and missing data, able to learn the dynamics of complex systems, and provides formal uncertainty quantification on parameter estimates and the confidence of the discovered dynamics.
Last, we apply our method to infer the evolution of atmospheric vorticity over time having only observed the streamfunction and obtain results that coincide with geophysical balances (i.e., the barotropic vorticity equation).

The remainder of this paper is organized as follows.
In Section \ref{sec:notation} we define the tensor and derivative notation used throughout the manuscript.
In Section \ref{sec:pdetection} we give background on the general dynamic system, showcase how inference on the derivative of the system is made, and present the Bayesian hierarchical model.
In Section \ref{sec:estimation} we describe parameter estimation and discuss modeling choices.
In Section \ref{sec:simulation} we demonstrate our method on multiple simulated data sets and in Section \ref{sec:barotrophic} we perform inference on a real-world system.
Section \ref{sec:conclusion} concludes the paper.

\section{Preliminary Notation}\label{sec:notation}

In this section we define tensor and derivative notation. 
All variables in this section are used only for illustrating notation.
Problem specific notation will be introduced in Section \ref{sec:pdetection}.

\subsection{Tensor Notation}\label{sec:tensor}

PDEs are commonly defined over multiple dimensions (e.g., space, time, components), and benefit from the use of higher-order tensor notation when the number of dimensions is three or more.
We generally follow the notation of \citet{Kolda2009} and refer the reader to their work for more details and references of tensor notation and applications.

Let $\EuScript{X} \in \mathbb{R}^{I_1 \times I_2 ... \times I_N}$ be a tensor of order $N$ where the $(i_1, i_2, ..., i_N)$ element is denoted by $\EuScript{X}(i_1, i_2, ..., i_N)$.
A \textit{slice} of the tensor is a two-dimensional section where all but two indices are held constant.
For example, the horizontal, lateral, and frontal slices of the third order tensor $\EuScript{Y} \in \mathbb{R}^{I \times J \times K}$ are denoted by $\vY_{i::}$, $\vY_{:j:}$, and $\vY_{::k}$, respectively.
A tensor can be converted to a matrix using $n$-mode matricization (also known as unfolding or flattening).
The $n$-mode matricization of the tensor $\EuScript{X}$, denoted by $\vX_{(n)}$, arranges the mode$-n$ fibers (the higher-order equivalent of matrix rows and columns) to be columns in the resulting matrix.
For example, the possible modes of $\EuScript{Y}$ are $\vY_{(1)} \in \mathbb{R}^{I \times (J \times K)}$, $\vY_{(2)} \in \mathbb{R}^{J \times (I \times K)}$, and $\vY_{(3)} \in \mathbb{R}^{K \times (I \times J)}$.
In general, we will only be concerned with the mode-3 matricization of a tensor and will denote $\vY$ in place of $\vY_{(3)}$ (all other modes will be properly denoted).

To multiply a tensor by a matrix $\vB \in \mathbb{R}^{I_n \times J}$, we use the $n$-mode product (i.e., multiply a tensor by a matrix or vector in mode $n$).
The $n$-mode product of the tensor $\EuScript{X}$ and matrix $\vB$ is denoted as $\EuScript{X} \times_n \vB$ and is of size $I_1 \times I_2 ... \times I_{n-1} \times J \times I_{n+1} \times ... \times I_N$.
Equivalently, in terms of unfolded (matricized) tensors, $\EuScript{Z} =  \EuScript{X} \times_n \vB$ $\Leftrightarrow$ $\vZ_{(n)} =  \vB \vX_{(n)}$.

\subsection{Tensor Basis Representation}\label{sec:tensor_basis}

Let $\EuScript{Y} \in \mathbb{R}^{I \times J \times K}$ be the order 3 tensor from Section \ref{sec:tensor}.
Define the expansion of the $(i,j,k)$ element of $\EuScript{Y}$ as
\begin{align*}
    y(i,j,k) \equiv \sum_{p=1}^\infty \sum_{q=1}^\infty \sum_{r=1}^\infty g(p,g,q) a(i,p) b(j,q) c(k,r),
\end{align*}
where $\{a(i,p): p = 1, ..., \infty\}$, $\{b(j,q): q = 1, ..., \infty\}$, and $\{c(k,r): r = 1, ..., \infty\}$ are basis functions and $\{g(p,g,q): p,q,r = 1, ..., \infty \}$ is the tensor of associated basis coefficients.
To reduce the dimension, we keep the first $P, Q$, and $R$ terms from $a, b$, and $c$, and define each basis function at discrete values $p = 1, ..., P$, $q = 1, ..., Q$, and $r = 1, ..., R$, respectively.
That is, let 
\begin{align*}
    \EuScript{Y} \approx \sum_{p=1}^P \sum_{q=1}^Q \sum_{r=1}^R g(p,q,r) \va(p) \circ \vb(q) \circ \vc(r)  = [\![ \EuScript{G}; \vA, \vB, \vC ]\!]  = \EuScript{G} \times_1 \vA \times_2 \vB \times_3 \vC,
\end{align*}
where $\circ$ is the vector outer product, $\vA$ is a $I \times P$, $\vB$ is a $J \times Q$, and $\vC$ is a $K \times R$ matrix of basis coefficients where each column is given by $\va(p) \equiv (a(1,p), ..., a(I,p))$, $\vb(q) \equiv (b(1,q), ..., a(J,q))$, and $\vc(r) \equiv (c(1,r), ..., c(K,r))$, and $[\![ \EuScript{G}; \vA, \vB, \vC ]\!]$ is shorthand notation introduced in \cite{Kolda2006}.
Our basis decomposition is similar to the Tucker decomposition \citep{Tucker1966}, except we assume $\vA, \vB,$ and $\vC$ are known and our goal is to estimate $\EuScript{G}$.
Note, we provide the expansion only for an order 3 tensor (sufficient for this manuscript), but the concept can be extended to higher order tensors.

\subsection{Derivative Notation}\label{sec:derivative_notation}

As discussed in Section \ref{sec:introduction}, we propose a method to discover the governing equations in PDEs.
As the name suggests, a PDE is composed of partial derivatives of some variable $u = u(x,y,t)$ that is indexed in space or time or both.
We denote partial derivatives using a subscript, for example $\frac{\partial u}{\partial t} = u_t$, $\frac{\partial u}{\partial x} = u_x$, $\frac{\partial^2 u}{\partial t^2} = u_{tt}$, and so forth.
We denote the $i$th order of a derivative generally as $\frac{\partial^{(i)} t}{\partial t^{(i)}} = u_{t^{(i)}}$.
In order to disambiguate notation, we denote the index of a vector/matrix/tensor using parentheses (e.g., $\va(i), \vA(i,j), \EuScript{A}(i,j,k)$), reserving the subscript to denote derivatives.

Within the PDE literature there are different choices of notation to denote the same operation.
For example, the Laplacian operator can be denoted as $\Delta u = \nabla^2 u = \nabla \cdot \nabla u = \frac{\partial^2 u}{\partial x^2} + \frac{\partial^2 u}{\partial y^2} = u_{xx} + u_{yy}$.
Wherever an operator such as the Laplacian is used for the first time, we will define it.
This may result in our notation differing from other texts, but we aim to be consistent within the paper.

\section{Bayesian Dynamic Equation Discovery}\label{sec:pdetection}

Here we propose a general hierarchical model for making inference on nonlinear spatio-temporal dynamic systems.
We begin by motivating the general class of PDEs and manipulate them to fit within a statistical framework.

\subsection{Dynamic Equations}\label{sec:dynamic_equations}

Consider the general PDE dynamic system describing the evolution of a continuous field $\{\vu(\vs, t): \vs \in D_s, t \in D_t\},$ 
\begin{align}\label{eqn:pde}
    \vu_{t^{(J)}}(\vs, t) = 
    M\left(\vu(\vs, t), \vu_x(\vs, t), \vu_y(\vs, t), \vu_{xy}(\vs, t), ..., \vu_{t^{(1)}}(\vs, t), ..., \vu_{t^{(J-1)}}(\vs, t), \vomega(\vs, t)
    \right)
\end{align}
where the vector $\vu(\vs, t) \in \mathbb{R}^N$ denotes the realization of the $N$-dimensional system at location $\vs$ and time $t$ (e.g., $\vu(\vs,t) = [u(\vs, t, 1), u(\vs, t, 2), ..., u(\vs, t, N)]'$), $M(\cdot)$ represents the (potentially nonlinear) evolution function, and $\vomega(\vs, t)$ represents any covariates that might be included in the system.
Here, $\vs \in \{\vs_1, ..., \vs_S\} = D_s$ is a spatial location in the domain with $|D_s| = S$, and $t \in \{1, ..., T\} = D_t$ is the temporal realization of the system where $|D_t| = T$.
We define (\ref{eqn:pde}) in two dimensions (i.e., $D_s \in \mathbb{R}^2$ and $\vs = (x,y)$), but the problem can be simplified to one dimension (i.e., $D_s \in \mathbb{R}^1$ and $\vs = x$) or generalized to higher spatial dimensions.
Finally, as is common in the dynamic systems literature, we refer to the $N$-dimensional multivariate vector $\vu(\vs,t)$ as the state or system state, and use the term \textit{component} to refer to each of the $N$ elements of $\vu(\vs,t)$.

We reparameterize \ref{eqn:pde} to be intrinsically linear (in parameters) as
\begin{align}\label{eqn:sparse_system}
    \vu_{t^{(J)}}(\vs, t) =
    \vM \vf\left(\vu(\vs, t), \vu_x(\vs, t), \vu_y(\vs, t), \vu_{xy}(\vs, t), ..., \vu_{t^{(1)}}(\vs, t), ..., \vu_{t^{(J-1)}}(\vs, t), \vomega(\vs, t)
    \right),
\end{align}
where $\vM$ is a $N \times D$ \textit{sparse} matrix of coefficients and $\vf(\cdot)$
is a vector-valued nonlinear transformation function of length $D$.
The input of the arguments for $\vf(\cdot)$ are general and contain anything that \textit{potentially} relates to the system.
For example, this could include terms describing advection, diffusion, dispersion and growth, polynomial functions and interactions, or sinusoidal functions, and are chosen based on a general mechanistic understanding of the system.
This results in $D$ being quite large and (\ref{eqn:sparse_system}) has the potential to be highly over-parameterized.
Thus, we will employ regularization to induce sparsity in the matrix $\vM$.

As an example of a classic PDE within our framework, consider the reaction diffusion equation
\begin{align*}
    \vu_t(\vs, t) = \vD \nabla^2\vu(\vs,t) + \vc(\vu(\vs,t)),
\end{align*}
where $\vu(\vs,t) = [b(\vs,t), d(\vs,t)]'$ represents the densities of two processes, $\vD$ is a diagonal matrix where $diag(\vD) = [D_b, D_d]$ are the diffusion constants, and $\vc(\vu(\vs,t)) = [c_b(b(\vs,t),d(\vs,t)), \\ c_d(b(\vs,t),d(\vs,t))]'$ are (non)linear reaction functions.
The reaction-diffusion equation can be used to model the densities of prey ($b$) and predator ($d$) populations \citep{Hastings1996}.
For a predator-prey model, a possible choice for the reaction function is $\vc(\vu(\vs,t)) = [\gamma b - \delta b d, - \mu d + \eta b d]'$, a simplistic representation of the Lotka-Volterra system where $\gamma$ and $\delta$ represent the prey's birth and predation rates, respectively, and $\mu$ and $\eta$ represent the predator death and kill success rates.
Following (\ref{eqn:sparse_system}), and suppressing the spatial and temporal indices, we have
\begin{align*}
    \begin{bmatrix}
        b_t\\
        d_t
    \end{bmatrix} = 
    \begin{bmatrix}
        \gamma & 0 & -\delta & D_b & D_b & 0 & 0  \\
        0 & -\mu & \eta & 0 & 0 & D_d & D_d 
    \end{bmatrix}
        \begin{bmatrix}
        b & d & bd & b_{xx} & b_{yy} & d_{xx} & d_{yy}
    \end{bmatrix}'.
\end{align*}
Typically we do not know $\vf(\cdot) = [b, d, bd, b_{xx}, b_{yy}, d_{xx}, d_{yy}]'$ and instead highly over-parameterize $\vf(\cdot)$ by including a library of potential terms and select against the coefficients in $\vM$ to identify relevant terms.

In real-world problems, (\ref{eqn:sparse_system}) does not hold exactly.
Stochastic forcing could perturb the system (e.g., weather systems, demographic stochasticity) or there could be error in the model specification.
We accommodate this unknown stochasticity by including an additive error term 
\begin{align}\label{eqn:sto_system}
    \vu_{t^{(J)}}(\vs, t) = \vM \vf\left(\vu(\vs, t), \vu_x(\vs, t), \vu_y(\vs, t), \vu_{xy}(\vs, t),..., \vu_{t^{(1)}}(\vs, t), ..., \vu_{t^{(J-1)}}(\vs, t), \vomega(\vs, t)
    \right) + \veta(\vs, t),
\end{align}
where, for example, $\veta(\vs, t) \overset{\text{i.i.d.}}{\sim} N(\vec{0}, \vSigma_U)$ is a mean zero Gaussian process with variance/covariance matrix $\vSigma_U$.
In general, spatial or temporal dependencies could be considered in this error term.

To represent (\ref{eqn:sto_system}) using tensor notation, let $\EuScript{U} = \{u(\vs, t, n): \vs \in D_s, t = 1, ..., T, n = 1, ..., N\}$ where $\EuScript{U} \in \mathbb{R}^{S \times T \times N}$ is the tensor of the dynamic process.
Similarly, let $\EuScript{F}\in \mathbb{R}^{S \times T \times D}$ be the function $\vf(\cdot)$ evaluated at each location in space-time and $\widetilde{\veta} \in \mathbb{R}^{S \times T \times N}$ is the space-time-component uncertainty tensor.
The tensor formulation of (\ref{eqn:sto_system}) in then
\begin{align}\label{eqn:process_tensor}
    \EuScript{U}_{t^{(J)}} = \EuScript{F} \times_3 \vM + \widetilde{\veta}.
\end{align}
This forms the core of our process model, where we relate the temporal derivative of some space-time-component process to a nonlinear function of its current state.
While not explicitly stated in (\ref{eqn:process_tensor}), $\EuScript{F}$ is still a function of the state process $\EuScript{U}$.

\subsection{Basis Representation}\label{sec:basis}

As described in Section \ref{sec:tensor_basis}, we can represent the $\EuScript{U}$ tensor using basis functions.
Decomposing $\EuScript{U}$ in terms of a finite collection of spatial, temporal, and component basis functions, we write
\begin{align*}
    \EuScript{U} & \approx \sum_{p=1}^P \sum_{q=1}^Q \sum_{r=1}^R a(p,q,r) \vpsi(p) \circ \vphi(q) \circ \vtheta(r) = \EuScript{A} \times_1 \vPsi \times_2 \vPhi \times_3 \vTheta \coloneqq [\![ \EuScript{A}; \vPsi, \vPhi, \vTheta ]\!]
\end{align*}
where $\EuScript{A} \in \mathbb{R}^{P \times Q \times R}$, $\vPsi \in \mathbb{R}^{S \times P}$, $\vPhi \in \mathbb{R}^{T \times Q}$, and $\vTheta \in \mathbb{R}^{N \times R}$.
Here, $\vPsi, \vPhi$, and $\vTheta$ are matrices of spatial, temporal, and component basis functions, respectively, and $\EuScript{A}$ is a tensor of basis coefficients (traditionally called the \textit{core tensor}).

We can obtain derivatives of the elements of $\EuScript{U}$ analytically by taking derivatives of the basis functions.
Specifically, let $\vPsi$ and $\vPhi$ be matrices of basis functions differentiable up to at least the highest order considered in (\ref{eqn:pde}).
We then compute spatial and temporal derivatives of $\EuScript{U}$ by computing the derivatives of $\vPsi$ and $\vPhi$.
That is, denote $\frac{\partial}{\partial x} \vPsi = \vPsi_x$, $\frac{\partial}{\partial y} \vPsi = \vPsi_y$, $\frac{\partial}{\partial t} \vPhi = \vPhi_t$, and so forth.
Derivatives of $\EuScript{U}$ are then computed as
\begin{align}\label{eqn:derivative_example}
\begin{split}
    \frac{\partial}{\partial t} \EuScript{U} & = \EuScript{A} \times_1 \vPsi \times_2 \vPhi_t \times_3 \vTheta = [\![ \EuScript{A}; \vPsi, \vPhi_{t}, \vTheta ]\!] \\
    \frac{\partial}{\partial x} \EuScript{U} & = \EuScript{A} \times_1 \vPsi_x \times_2 \vPhi \times_3 \vTheta = [\![ \EuScript{A}; \vPsi_{x}, \vPhi, \vTheta ]\!] \\
    \frac{\partial^2}{\partial x \partial y} \EuScript{U} & = \EuScript{A} \times_1 \vPsi_{xy} \times_2 \vPhi \times_3 \vTheta = [\![ \EuScript{A}; \vPsi_{xy}, \vPhi, \vTheta ]\!],
\end{split}
\end{align}
and so forth.
Representing (\ref{eqn:process_tensor}) using the basis decomposition, we have
\begin{align*}
    [\![ \EuScript{A}; \vPsi, \vPhi_{t^{(J)}}, \vTheta ]\!] = \EuScript{F} \times_3 \vM + \veta,
\end{align*}
where $\veta$ may include truncation error.
While not explicitly stated, $\EuScript{F}$ now depends on $\vPsi, \vPhi, \vTheta$, and $\EuScript{A}$.

\begin{proposition}\label{prop:prop1}
The mode-3 decomposition of $[\![ \EuScript{A}; \vPsi, \vPhi_{t^{(J)}}, \vTheta ]\!] = \EuScript{F} \times_3 \vM + \widetilde{\veta}$ where $\veta(\vs,t) \overset{i.i.d.}{\sim} N_N(\vec{0}, \Sigma_U)$ in space and time at location $\vs$ and time $t$ is
\begin{align*}
    \vTheta \vA (\vphi_{t^{(J)}}(t) \otimes \vpsi(\vs))' = \vM \vf(\vA, \vpsi(\vs), \vpsi_x(\vs), \vpsi_y(\vs), \vpsi_{xy}(\vs), ..., \vphi_{t^{(0)}}(t), ..., \vphi_{t^{(J)}}(t), \vomega(\vs,t)) + \veta(\vs,t),
\end{align*}
where $\vA$ is a $R \times PQ$ matrix of basis coefficients, $\vpsi(\vs)$ is a length-$P$ vector of spatial basis functions, $\vphi(t)$ is a length-$Q$ vector of temporal basis functions, and $\vTheta$ is a $N \times R$ matrix of component basis functions.
\end{proposition}

\noindent \textit{Proof.} See supplementary material.

Decomposing (\ref{eqn:process_tensor}) in terms of basis functions and taking the mode-3 matricization accomplishes two tasks.
First, this enables inference on derivatives of the process $\vu(\vs, t)$ when only the process is known (e.g., see (\ref{eqn:derivative_example})).
Second, keeping fewer basis functions than observations (e.g., $P < S$, $Q < T$) allows the reconstruction of $\EuScript{U}$ to be smooth \citep{Wang2016}.

Note, we include $\vTheta$ for generality in the construction of our method.
While one could specify $\vTheta$ in terms of basis functions, our goal is not to reduce the dimension of the system state variables.
In our analyses, we choose $\vTheta$ to be the identity matrix.

\subsection{Transformation of Derivative}\label{sec:transform}

Up to this point, we have considered PDEs that relate the temporal derivative (of some order $J$) of the continuous surface $\vu$ on left-hand side (LHS) of (\ref{eqn:pde}) to a function of its current state on the right-hand side (RHS) of (\ref{eqn:pde}).
However, equations with a spatio-temporal derivative of $\vu$ on the LHS are common \citep[e.g., vorticity equation,][]{Higham2016a}.
For example, the LHS of (\ref{eqn:pde}) could depend on the Laplacian operator, where $\nabla^2\vu_{t^{(J)}}(\vs,t) = \vu_{xxt^{(J)}}(\vs,t) + \vu_{yyt^{(J)}}(\vs,t)$.

To be more general, we now allow the LHS of (\ref{eqn:pde}) to be a function of spatio-temporal derivatives of $\vu$ and consider the more general PDE
\begin{align}\label{eqn:gen_pde}
    g(\vu_{t^{(J)}}(\vs, t)) = 
    M\left(\vu(\vs, t), \vu_x(\vs, t), \vu_y(\vs, t), \vpsi_{xy}(\vs), ..., \vu_{t^{(1)}}(\vs, t), ..., \vu_{t^{(J-1)}}(\vs, t), \vomega(\vs, t)
    \right),
\end{align}
where $g(\cdot)$ is some linear differential operator.
The original PDE (\ref{eqn:pde}) is a special case of (\ref{eqn:gen_pde}) where $g(\cdot)$ is the identity function.

\begin{proposition}\label{prop:prop2}
Let $g(\cdot)$ be a linear differential operator.
The basis formulation of a PDE with a space-time function $g(\vu_{t^{(J)}}(\vs, t))$ on the LHS is 
\begin{align*}
    \vTheta \vA (\vphi_{t^{(J)}}(t) \otimes g(\vpsi(\vs)))'.
\end{align*}
\end{proposition}

\noindent \textit{Proof.} See supplementary material.

From Proposition \ref{prop:prop2}, the basis representation of a PDE with a spatio-temporal function on the LHS is
\begin{align}\label{eqn:gen_process_model}
\begin{split}
    \vTheta \vA (\vphi_{t^{(J)}}(t) \otimes g(\vpsi(\vs)))' & = \\
    \vM \vf(\vA, \vpsi(\vs), \vpsi_x(\vs) & , \vpsi_y(\vs), \vpsi_{xy}(\vs), ..., \vphi_{t^{(0)}}(t), ..., \vphi_{t^{(J)}}(t), \vomega(\vs,t)) + \veta(\vs,t),
\end{split}
\end{align}
where $\veta(\vs, t) \overset{i.i.d.}{\sim} N_N(\vec{0}, \Sigma_U)$ in space and time.
Completing the example from before using the $g = \nabla^2$ Laplacian operator, the LHS for (\ref{eqn:gen_process_model}) is $\vTheta \vA (\vphi_{t^{(J)}}(t) \otimes  (\vpsi_{xx}(\vs) + \vpsi_{yy}(\vs)))'$.

\subsection{Data Model}\label{sec:data_model}

We assume $\vv(\vs, t)$ is an observation of the $N$-dimensional latent process outlined in Section \ref{sec:basis} with some unknown measurement uncertainty.
We model $\vv(\vs, t)$ using a generalization to the traditional linear data error model that links the dynamics to the observed process \citep[e.g., see][Chapter~7]{Cres}.
That is, we model
\begin{align}\label{eqn:base_data_model}
    \vv(\vs, t) = \vH(\vs,t) \vu(\vs, t) + \widetilde{\vepsilon}(\vs, t),
\end{align}
where $\vv(\vs, t) \in \mathbb{R}^{L(\vs,t)}$, $\vH(\vs,t) \in \mathbb{R}^{L(\vs,t) \times N}$ is the incidence matrix that maps from $\vu(\vs,t)$ to $\vv(\vs, t)$, and uncertainty in the observations of the process are captured by $\widetilde{\vepsilon}(\vs,t) \overset{\text{indep.}}{\sim} N_{L(\vs,t)}(\vec{0}, \widetilde{\vSigma}_V(\vs,t))$.
The dimension of the data, $L(\vs,t)$, is allowed to vary based on the space-time location due to potentially missing data and we assume the errors are independent in space and time.

Within the hierarchical model, missing data are accommodated by allowing the dimension of the incidence matrix, $\vH(\vs, t)$, to vary in time.
Since missing data are handled in the data model and the latent process is fully specified, missing data do not impact the process model specification.
If there are no missing data at time $t$ and location $\vs$, then $L(\vs,t) = N$ and $\vH(\vs, t) = \vI_N$.
When one or more system components are missing data, the row corresponding to the missing system component is removed.
For example, if we have a three-dimensional system, say $\vu(\vs, t) = [a(\vs, t), b(\vs, t), c(\vs, t)]$ and the observation component for $b(\vs, t)$ is missing at location $\vs$ and time $t$, then 
\begin{align*}
    \vH(\vs, t) = \begin{bmatrix} 1 & 0 & 0 \\ 0 & 0 & 1 \end{bmatrix}.
\end{align*}
See Chapter 7 of \cite{Cres} for more discussion of this approach for accommodating missing observations in hierarchical spatio-temporal models.

Incorporating the basis expansion of the process in (\ref{eqn:base_data_model}), at location $\vs$ and time $t$,
\begin{align}\label{eqn:data_model}
    \vv(\vs,t) = \vH(\vs,t) \vTheta \vA (\vphi_{t^{(0)}}(t) \otimes \vpsi(\vs))' + \vepsilon,
\end{align}
where $\vepsilon(\vs,t) \overset{indep.}{\sim} N_L(\vec{0}, \Sigma_V(\vs,t))$ and $\vepsilon(\vs, t)$ now accounts for the discrepancy between the ``true'' underlying process and our approximation using the basis formulation.

\subsection{Parameter Model}\label{sec:parameter_model}

The data and process equations correspond to the first and second level of our hierarchical model, respectively.
For convenience, we restate (\ref{eqn:data_model}) and (\ref{eqn:gen_process_model}) for location $\vs$ and time $t$
\begin{align*}
    \vv(\vs, t) & = \vH(\vs,t) \vTheta \vA (\vphi_{t^{(0)}}(t) \otimes \vpsi(\vs))' + \vepsilon(\vs, t) \\
    \vTheta \vA (\vphi_{t^{(J)}}(t) \otimes g(\vpsi(\vs)))' & = \vM \vf(\vA, \vpsi(\vs), \vpsi_x(\vs), \vpsi_y(\vs), \vpsi_{xy}(\vs), ..., \vphi_{t^{(0)}}(t), ..., \vphi_{t^{(J)}}(t), \vomega(\vs,t)) + \veta(\vs, t),
\end{align*}
where $\vepsilon(\vs, t) \overset{indep.}{\sim} N_{L(\vs,t)}(\vec{0}, \vSigma_V(\vs, t))$ and $\veta(\vs, t) \overset{i.i.d.}{\sim} N_N(\vec{0}, \vSigma_U)$.
For clarity, we present the details on the model parameters in Table \ref{tab:paramerers}.
Our goal is to make inference on the unknown parameters $\vM, \vSigma_U, \vSigma_V(\vs, t)$, and $\vA$.
The sparse matrix $\vM$ identifies the nonlinear dynamic equation, $\vSigma_U$ captures the error dependencies within the dynamic equation, $\vSigma_V(\vs, t)$ captures the measurement uncertainty associated with the observed process, and $\vA$ defines the smooth latent process.

\begin{table}[ht!]
    \centering
    \begin{tabular}{|c|l|l|l|}
        \hline
        Model       & Symbol       & Description              & Dimension         \\
        \hline 
                    & Variable            &                          &                   \\
        Data        & $ \vv(\vs,t) $      & Observed data            & $ L(\vs,t) \times 1 $  \\
        Data        & $ \vH(\vs,t) $      & Mapping matrix           & $ L(\vs,t) \times N $  \\
        Data        & $ \vepsilon(\vs,t)$ & Data uncertainty distribution & $L(\vs,t) \times 1 $  \\
        Data        & $ \vSigma_V(\vs,t)$ & Measurement error covariance matrix  & $L(\vs,t) \times L(\vs,t)$ \\
        Process     & $ \vu(\vs, t) $     & Dynamic process          & $ N \times 1 $    \\
        Process     & $ \EuScript{A} $    & Basis coefficient tensor       & $ P \times Q \times N $  \\
        Process     & $ \vA $             & Basis coefficient matrix (mode-3)       & $ N \times (P \times Q) $  \\
        Process     & $ \vpsi(\vs) $      & spatial basis function for location $\vs$ & $ P \times 1 $  \\
        Process     & $\vphi_{t^{(j)}}(t)$& $j$th order temporal basis function for time $t$ & $ Q \times 1 $  \\
        Process     & $ \vTheta $         & component basis function matrix & $ N \times R $  \\
        Process     & $ \vM $             & Dynamic evolution matrix & $ N \times D $    \\
        Process     & $ \vf(\cdot) $      & Feature library          & $ D \times 1 $    \\
        Process     & $ \veta(\vs,t) $    & Process uncertainty distribution      & $ N \times 1 $    \\
        Process     & $ \vSigma_U $       & Dynamic equation error covariance matrix  & $N \times N$ \\ 
        \hline
                    & Dimension           &                          &                   \\
                    & $ T $               & Number of observed time points                 & 1 \\
                    & $ S $               & Number of observed spatial locations           & 1 \\
                    & $ L(\vs,t) $        & Dimension of observation vector at time $t$ and location $\vs$ & 1 \\
                    & $ N $               & Dimension of latent process (dynamic system)   & 1 \\
                    & $ D $               & Number of library functions                    & 1 \\
                    & $ P $               & Number of spatial basis functions              & 1 \\
                    & $ Q $               & Number of temporal basis functions             & 1 \\
                    & $ R $               & Number of component basis functions            & 1 \\
                    & $ J $               & Highest order derivative in the dynamic system & 1 \\ 
        \hline
                    & Indices             &                          &                   \\
                    & $ t $               & Time interval, $t \in \{1, ..., T\} = D_t$, $|D_t| = T$   & 1 \\
                    & $ \vs $             & Spatial location, $\vs \in \{\vs_1, ..., \vs_S\} = D_s$, $|D_s| = S$ & 1 \\
                    & $ j $               & Order of the derivative, $j = 1, ..., J$   & 1 \\
        \hline
    \end{tabular}
    \caption{List of symbols used in the Bayesian hierarchical model.}
    \label{tab:paramerers}
\end{table}

To complete our Bayesian hierarchical model, we define the following priors on these parameters.
We use the spike-and-slab prior \citep{Mitchell1988, George1993} to induce sparsity into $\vM$.
We write
\begin{align*}
    \vM(n)|\vgamma(n), \vsigma^2_U(n) = \prod_{d=1}^{D} [(1-\vgamma(n,d))\delta_0 + \vgamma(n,d) p(M(n,d)|\vsigma^2_U(n), \cdot)],
\end{align*}
where $M(n,d)$ denotes coefficient $d$ of component $n$, $\vgamma$ is a matrix of inclusion indicators of the same dimension as $\vM$, $\delta_0$ denotes the Dirac function at 0, $\vsigma^2_U(n)$ is the $n$th diagonal component of $\vSigma_U$, $p(\vgamma(n,d) = 1|\pi_n) = \pi(n)$, and $\pi(n) \sim Beta(a, b)$.
That is, if a variable is not included (i.e., $\vgamma(n,d) = 0$), then the corresponding element $M(n,d)$ is zero.
If a variable is included (i.e., $\vgamma(n,d) = 1$), then the corresponding element $M(n,d)$ is non-zero.
There are multiple choices for the prior $p(M(n,d)|\vsigma^2_U(n), \cdot)$.
We specify the g-slab prior corresponding to Zellner's g-prior \citep{Zellner1986a} where $g$ is taken to be the size of the data.
See \citet{Malsiner-Walli2016} for other potential choices and further discussion. 

While other shrinkage/selection priors could be used, such as Stochastic Search Variable Selection \citep[SSVS;][]{George1993}, LASSO \citep{Park2008}, or Horseshoe \citep{Carvalho2010}, we found the spike-and-slab to be preferable since it performs well with correlated predictors \citep{Roskova2014}, which is generally present in the feature library (see Section \ref{sec:estimation}).
Additionally, the posterior summary of the latent variable $\vgamma(n,d)$ gives the inclusion probability for each component of $\vM$, providing further insight into the certainty of the recovered system.
For all examples presented below, we determine the identified system as composed of terms that are included with at least 50\% posterior probability.
However, this threshold is subjective and one could choose a different value depending on their specific application.

To estimate $\vgamma$ and avoid reducibility of the Markov chain, we compute the marginal posterior distribution $[\vgamma|\vA, \vTheta, \vPhi_{t^{(J)}}, \vPsi, \vSigma_U] \propto [\vA, \vTheta, \vPhi_{t^{(J)}}, \vPsi|\vgamma, \vSigma_U][\vgamma]$, which is obtained by integrating over the parameters subject to selection.
That is,
\begin{align*}
    [\vA, \vTheta, \vPhi_{t^{(J)}}, \vPsi|\vgamma, \vSigma_U] = \int \int [\vA, \vTheta, \vPhi_{t^{(J)}}, \vPsi|\vM, \vSigma_U, \vgamma][\vM][\vSigma_U] d\vM d\vSigma_U.
\end{align*}
To make the integration analytically tractable and keep conjugacy in the model, we restrict $\vSigma_U$ to be diagonally structured where $\vSigma_U = diag(\sigma^2_{U}(1), ..., \sigma^2_{U}(N))$ and each diagonal element is assigned the non-informative prior $\sigma^2_{U}(n) \propto 1/\sigma^2_{U}(n), n = 1, ..., N$.
Then, the probability any element is included is given as
\begin{align}\label{eqn:delta_prob}
    p(\vgamma(n,d)=1|\cdot) = \frac{1}{1 + \frac{1-\pi(n)}{\pi(n)}R_{\gamma}(n,d)}
\end{align}
where
\begin{align*}
    R_{\gamma}(n,d) = \frac{[\vA, \vTheta, \vPhi_{t^{(J)}}, \vPsi|\vgamma(n,d) = 0]}{[\vA, \vTheta, \vPhi_{t^{(J)}}, \vPsi|\vgamma(n,d) = 1]}.
\end{align*}
In situations where dependence between the components is required, a different prior could be used.

There is potential for elements of the variance-covariance matrix $\vSigma_V(\vs, t)$ to have small values.
Inference using traditional conjugate Inverse Gamma/Wishart priors are overly sensitive to the choice of hyperpriors when estimates are small \citep{Gelman2006}.
Instead, we use the conjugate Half-t prior proposed by \cite{Huang2013} for covariance estimation, which imposes less prior information and does not have as strong of influence on small estimates.
We restrict the measurement error to be diagonally structured since it is often a reasonable assumption that measurement noise is independent \citep{Cres} (although this restriction can be removed if warranted).
Let $\vSigma_V(\vs, t) = \vH(\vs, t) diag(\sigma^2_{V}(1), ..., \sigma^2_{V}(N)) \vH(\vs, t)'$, where each diagonal element, $\sigma^2_{V}(1), ..., \sigma^2_{V}(N)$, is assigned a conjugate Half-t$(2, 10^{-5})$ prior.

Finally, in order to induce sparcity in the basis coefficients, we assign a Bayesian elastic net prior \citep{Li2010} to $\vA$.
Specifically, our prior is 
\begin{align*}
    \pi(\vA) \propto \exp\{-\lambda_1\|\vA\|_1 -\lambda_2\|\vA\|_2^2\},
\end{align*}
where $\lambda_1, \lambda_2$ are penalty parameters.
The elastic net prior helps regularize the basis coefficients against basis functions.
While it is possible to specify hyperpriors for the two penalty terms, we find inference is not overly sensitive to the choice of penalty parameters and fix them each to a small value (e.g., 0.01 or 0.001).

\section{Model Estimation}\label{sec:estimation}

Our goal is to obtain samples from the joint posterior distribution $[\vM, \vSigma_U, \vSigma_V, \vgamma, \vA|\cdot]$.
We achieve this by sampling from the five full-conditional distributions $[\vM|\cdot]$, $[\vSigma_U|\cdot]$, $[\vSigma_V|\cdot]$, $[\vgamma|\cdot]$, and $[\vA|\cdot]$ (see the supplementary material for the details of the distributions and sampling algorithm) using a Markov chain Monte Carlo (MCMC) sampling scheme.
The four components $\vM, \vSigma_U, \vSigma_V$ and $\vgamma$ are updated using classical Bayesian methods and $\vA$ is updated using a stochastic gradient approach.
Due to the variety of problems for which our method is applicable, some modeling choices are case specific.
Additionally, some aspects of the implementation of the MCMC framework warrant a more detailed discussion.
The following sections provide additional information pertaining to these model specifications and procedures.

\subsection{Basis Coefficient Estimation}\label{sec:basis_estimation}

The basis coefficients, $\vA$, completely define the latent process and all derivatives in both space and time, meaning proper estimation is crucial to the discovery process.
Since $\vA$ is embedded within the nonlinear function $\vf(\cdot)$ (see Proposition \ref{prop:prop1}) and $f(\cdot)$ is problem specific, it is difficult to estimate.
To accommodate a generically specified $\vf(\cdot)$, we use an adapted version of stochastic gradient descent (SGD) with a constant learning rate \citep[SGDCL;][]{Mandt2016a}.
Whereas other approaches to estimate $\vA$ (e.g., Expectation-Maximization or Metropolis-Hastings) could be used, SGDCL provides important advantages -- a conjugate updating scheme and a reduced computational cost for any specification of $\vf(\cdot)$.

As with SGD, SGDCL relies on the gradient of a loss function and a learning rate.
For SGDCL, the loss function is the negative log posterior for our parameters of interest, $\vA$.
The loss function at location $\vs$ and time $t$ is
\begin{align*}
    \mathcal{L}(\vA; \vs, t) = - \log(&[\vv(\vs,t)| \vA, \vH(\vs,t), \vTheta, \vphi_{t^{(0)}}(t), \vpsi(\vs), \vSigma_V] \\
    & [\vA, \vTheta, \vphi_{t^{(i)}}(t), g(\vpsi(s))| \vM, \vSigma_U, \vA, \vTheta, \vphi_{t^{(0)}}(t), ..., \vphi_{t^{(J-1)}}(t), \vpsi(\vs), \vpsi_x(\vs), ...]) \\
    - \log(&[\vA]).
\end{align*}
To simplify notation, denote $\vB_0(\vs, t) = \vphi_{t^{(0)}}(t) \otimes \vpsi(\vs)$ and $\vB_J(\vs, t) = \vphi_{t^{(J)}}(t) \otimes g(\vpsi(\vs))$.
Then, the gradient of the loss function $\mathcal{L}(\vA; \vs, t)$, $\frac{\partial \mathcal{L}(\vA; \vs, t)}{\partial \vA} = \nabla_{\vA} \mathcal{L}(\vA; \vs, t)$ for location $\vs$ and time $t$ is
\begin{align*}
    \nabla \mathcal{L}(\vA; \vs, t) = & 
    -\vTheta' \vH'(\vs, t) \vSigma_V^{-1} \vv(\vs, t) B_0(\vs, t) + 
    \vTheta' \vH'(\vs, t) \vSigma_V^{-1} \vH(\vs, t) \vTheta \vA \vB'_0(\vs,t) \vB_0(\vs,t) \\
    & + \vTheta' \vSigma_U^{-1} \vTheta \vA \vB'_J(\vs,t) \vB_J(\vs,t) 
    - \vTheta' \vSigma_U^{-1} \vM \vf(\vs,t) \vB_J(\vs,t) \\
    & - \vB_J(\vs,t) \vA' \vTheta' \vSigma_U^{-1} \vM \bdot{\vF}'(\vs,t)
    + \vf'(\vs,t) \vM' \vSigma_U^{-1} \vM \bdot{\vF}'(\vs,t) \\
    & + \frac{1}{ST}\left( \lambda_1 sign(\vA) + 2 \lambda_2 \vA \right),
\end{align*}
where $\bdot{\vF}(\vs,t)$ generically denotes $\frac{\partial}{\partial \vA} \vf(\vA, \cdot)$.

SGDCL \citep{Mandt2016a} replaces the true gradient with the stochastic estimate,
\begin{align*}
    \widehat{\nabla \mathcal{L}}_{\mathcal{Z}}(\vA) = \frac{1}{|\mathcal{Z}|} \sum_{z \in \mathcal{Z}} \nabla_{\vA} \mathcal{L}(\vA; z),
\end{align*}
where $\mathcal{Z} \subset D_s \times D_t$ is a random subset of the observations, called a mini-batch, and $|\mathcal{Z}|$ is the cardinality of the set.
Within the context of a MCMC algorithm, the $\ell$th update of $\vA$ is given by
\begin{align*}
    \vA^{(\ell)} = \vA^{(\ell-1)} - \kappa \widehat{\nabla \mathcal{L}}_{\mathcal{Z}^{(\ell)}}(\vA^{(\ell-1)}),
\end{align*}
where $\mathcal{Z}^{(\ell)}$ denotes a random minibatch specific to the $\ell$ update and $\kappa$ is the learning rate.
To accommodate different scales for each component, we allow $\kappa$ to be a vector of length $N$ where each component can have a specific learning rate.

The final challenge to estimating $\vA$ is computing $\bdot{\vF}(\vs,t)$.
Because $\vf(\cdot)$ is problem specific, $\bdot{\vF}(\vs,t)$ is also problem specific.
One option is to use automatic differentiation (AD) to analytically compute the derivative of $\vf(\cdot)$.
There are many different libraries and programs that perform AD, and we explored the use of the \textit{ForwardDiff} \citep{Revels2016} package in Julia \citep{Bezanson2017} with success.
However, there is computational overhead to AD.
For all the examples presented here we computed $\bdot{\vF}(\vs,t)$ without AD for each problem to mitigate this computation bottleneck.

\subsection{Choice of Basis Functions}\label{sec:basis_choise}

The choice of basis functions are subjective and have the potential to affect the model fit (North et al. 2022).
Furthermore, the choices of spatial and temporal basis functions do not need to be the same (e.g., radial basis functions in space and Fourier basis functions in time).
There are other choices regarding basis functions that need to be taken into consideration \citep[see][Chapter~3 for a discussion on how to choose basis functions based on the ``shape'' of the data]{Ramsay2005}.
The most important requirement is that the spatial and temporal basis functions need to be differentiable up to at least the highest order spatial and temporal derivative considered, respectively.
We found local basis functions (e.g., B-splines) perform better than global basis functions (e.g., Fourier basis functions) \citep[see][]{North2022}, especially when there are local regions with minimal curvature.
For these reasons, we use B-splines of an order greater than our highest derivative (in both space and time).
In choosing how many basis functions to use, enough need to be included such that the estimated solution curve is flexible, the dynamics are captured, and the posterior latent space is properly explored, but not so many such that unnecessary noise is introduced into the system.
Empirically, we found a ratio of approximately 1 basis function to every 3 to 5 observations to work well.

\subsection{Choice of Feature Library}\label{sec:library_choice}

The choice of functions for the feature library is crucial to the identification of the system.
Our method is restricted to search over a predefined set of functions, meaning that our method is unable to identify an important function if it is not included in the library.
For this reason, it is best to over-parameterize the feature library (and hence $\vM$) instead of specifying a restrictive set of functions.
Additionally, some knowledge of the problem is beneficial (i.e., this is not a black-box approach).
Having an understanding of the potential dynamics \textit{a priori} can assist in the recovery of important dynamics.
For example, if the system appears to diffuse over time, then a diffusion term should be included.
A good default choice is polynomial terms that interact with varying orders of spatial and/or temporal derivatives of the process (e.g., see the library for Burgers' example) as this will cover a wide collection of systems.

With regard to the choice of $g()$ in (\ref{eqn:gen_pde}) and (\ref{eqn:gen_process_model}), scientific knowledge of the problem is required.
The choice of $g()$ is not searched over as with the library terms; rather it is pre-specified.
For example, in our real-world example, the  \textit{a prior} goal was to make inference on the change of atmospheric vorticity with time, and vorticity can be represented as the Laplacian of the streamfunction. 
Because this transformation function is not learned, it is a modeling choice that is user specified (the default is the identity as in Proposition \ref{prop:prop1}).

\subsection{Multicollinearity in Library}\label{sec:multicollinearity}

\begin{figure}[ht]
    \centering
    \includegraphics[width = 0.8\linewidth]{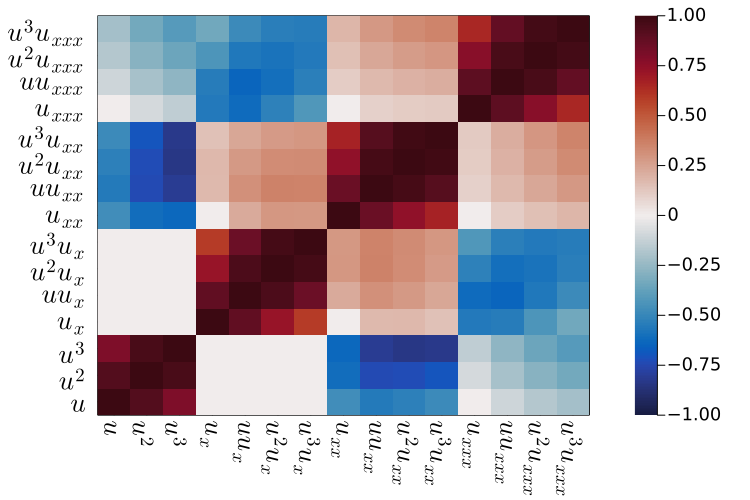}
    \caption{Correlation between terms of a potential feature library using data generated from Burgers' equation.}
    \label{fig:burger_corr}
\end{figure}

A major issue facing the identification of spatio-temporal dynamic equations is multicollinearity in the feature library.
Figure \ref{fig:burger_corr} shows the correlation between different components of a library using data generated from Burgers' equation.
In this example, the polynomial terms, $u, u^2,$ and $u^3$, are very positively linearly correlated $(\rho > 0.8)$, posing a challenge to parameter inference.
As with classical regression, multicollinearity has the potential to introduce bias into the coefficient estimates, including altering their sign.
While the spike-and-slab has been shown to perform well with correlated variables, as discussed in Section \ref{sec:parameter_model}, the problem still persists and can pose an estimation issue in problems similar to the example using Burgers' equation.

This issue originates from the over-inflation of minor reductions in the residual sum of squares (RSS) when a highly correlated, but incorrect term, is included.
Specifically, the probability of including any term in the library   (\ref{eqn:delta_prob}) is dependent on the ratio of the residual sum of squares for the model with the $M(n,d)$ term included ($RSS_{\gamma}$) to the model without the $M(n,d)$ term included ($RSS_{\setminus\gamma}$) through the value of $R_{\gamma}(n,d)$.
That is, under the $g$-prior
\begin{align*}
    R_{\gamma}(n,d) = (g+1)^{1/2}(RSS_{\gamma}/RSS_{\setminus\gamma})^{ST/2-1}.
\end{align*}
Because $RSS_{\gamma} < RSS_{\setminus\gamma}$, the ratio is bound between 0 and 1, where correct terms in the library result in the ratio being closer to 0 and incorrect terms result in the ratio being close to 1.
However, as the ratio is raised to a power of $ST/2 -1$ (proportional to the number of observations), the value of $R$ goes to 0 as the number of observations goes to infinity, resulting in all variables being found significant.
This issue is exacerbated by correlated variables, especially if there are multiple confounding variables where the system can be approximated by some linear combination of the feature library \textit{without} the true terms being included.

To combat this issue we propose a method to reduce the impact of correlated variables (i.e., variables where in the ratio of RSSs being close to 1 are found significant).
For this, we subsample the process when estimating the inclusion latent variable $\vgamma$.
That is, to compute (\ref{eqn:delta_prob}) within each iteration of the Gibbs sampler, we randomly sample the process.
This results in
\begin{align*}
    R^*_{\gamma}(n,d) = (g+1)^{1/2}(RSS_{\setminus\gamma}/RSS_{\gamma})^{S^*T^*/2-1},
\end{align*}
where $S^*$ and $T^*$ are the size of the subsampled dimensions.
We provide details on how to choose the subsample size and our choices for the examples in the supplementary material.
Note that this subsampling is only done for the $\vgamma$ update step of the algorithm.

\section{Simulations}\label{sec:simulation}

We show our proposed model is able to discover dynamic equations using data simulated from three well known systems -- Burgers' equation, the heat equation, and a reaction-diffusion system.
For all three examples, we investigate the impact of measurement noise on inference.
We simulate measurement noise by adding mean zero Gaussian errors to the state vector.
Specifically, we let $\vv(\vs, t) = \vu(\vs, t) + \zeta\vepsilon(\vs, t)$, where $\vu(\vs, t)$ is the simulated data, $ \vepsilon(\vs, t) \sim N(\mathbf{0}, \sigma^2\mathbf{I}_N)$ is the additive noise, $\sigma$ is the standard deviation of the simulated process $\vu(\vs, t)$, and $\zeta$ is the percent of noise ranging from 0 to 1.
In addition, we show how the model performs when data are missing sporadically for Burger's equation.
Unless otherwise stated, all reported estimates are rounded to three significant digits for readability.
For all simulations and real-world examples, we obtain 5000 posterior samples and discard the first 2500 as burn-in.
Convergence is assessed visually via trace plots with no issues detected.

\subsection{Burgers' Equation}

\begin{figure}[ht]
    \centering
    \includegraphics[width = 0.8\linewidth]{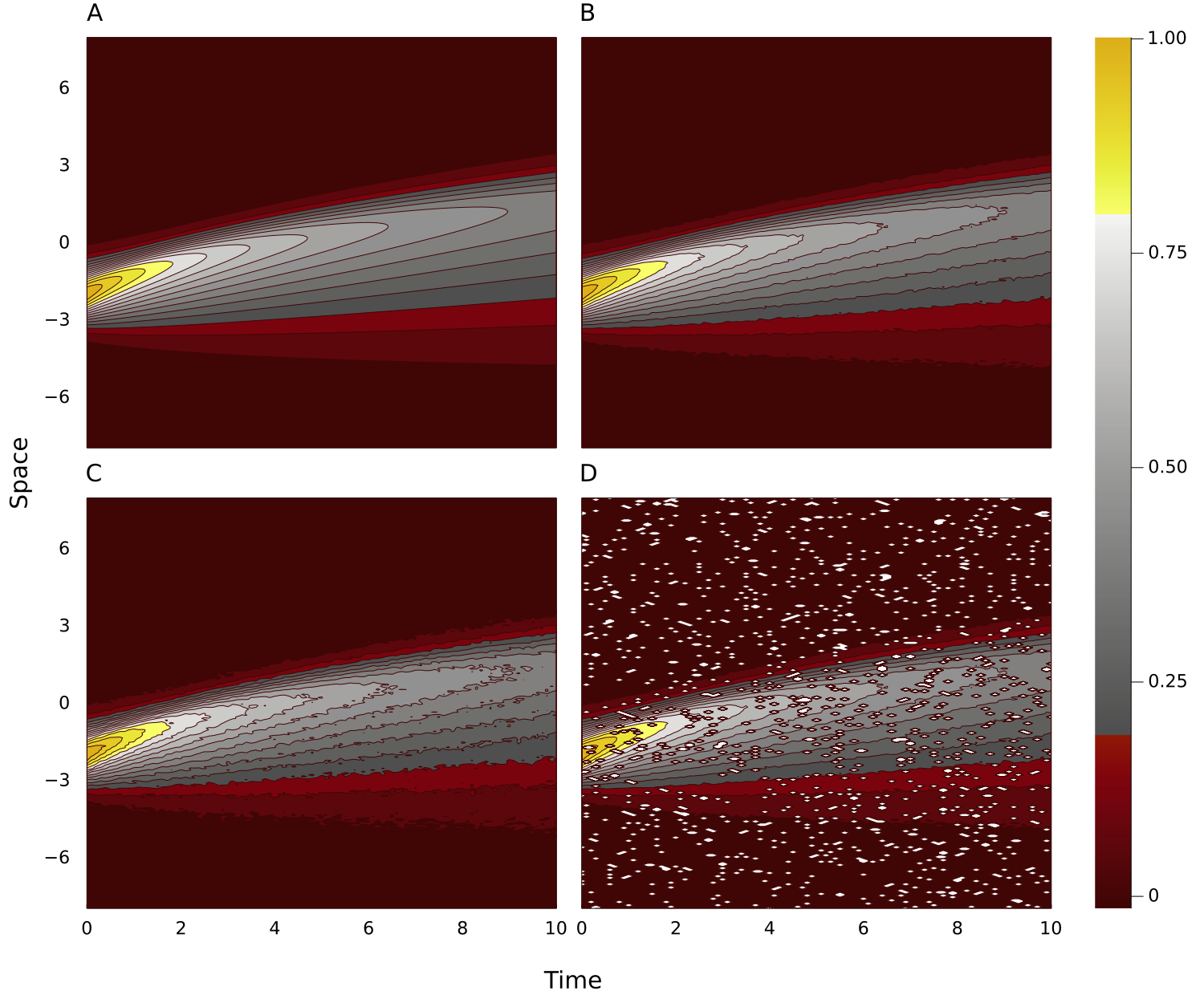}
    \caption{Data simulated from Burgers' equation with (A) no added measurement noise, (B) 2\% added measurement noise, (C) 5\% added measurement noise, and (D) 5\% of data missing at random and 2\% measurement noise.}
    \label{fig:burgers_data}
\end{figure}

Burgers' equation is a simplification of the Navier-Stokes equations, describing the speed of a fluid at a location in space and time \citep{Bateman1915, Burgers1948}.
We consider Burgers' equation in one spatial dimension defined by the nonlinear PDE
\begin{align*}
    u_t(x,t) = -u(x,t)u_{x}(x,t) + \nu u_{xx}(x,t),
\end{align*}
where $u(x,t)$ is the speed of the fluid at location $x$ and time $t$ and $\nu$ is the viscosity of the fluid.
Data are generated using spectral differentiation and the \textit{Tsit5} \citep{Tsitouras2011} numerical solver from the Julia package \textit{DifferentialEquations.jl} \citep{Rackauckas2017} with initial condition $u(x,0) = exp\{-(x+2)^2\}$.
The simulated data consist of 256 spatial locations across 101 time points where $D_s = [-8, 8]$ and $D_t = [0, 10]$ (Figure \ref{fig:burgers_data}).
We consider four cases -- no measurement noise, 2\% measurement noise, 5\% measurement noise, and 2\% measurement noise with 5\% of data missing at random.

For all four cases we specify the model with $P = 50, Q = 20, |\mathcal{Z}| = 100, \kappa = 10^{-4}$ and define the feature library as
\begin{align*}
    [u, u^2, u^3, u_x, u u_x, u^2 u_x, u^3 u_x, u_{xx}, u u_{xx}, u^2 u_{xx}, u^3 u_{xx}, u_{xxx}, u u_{xxx}, u^2 u_{xxx}, u^3 u_{xxx}].
\end{align*}
After obtaining posterior samples, we keep only terms with greater than a 50\% inclusion probability to be included in the identified equation.
The recovered equations and 95\% highest posterior density (HPD) interval without and with measurement noise for the included terms are shown in Table \ref{tab:burgers}.
In all four scenarios, the true components of the dynamic system are correctly identified.
In addition, the probability of including extraneous terms from the feature library is low in each scenario (see Supplementary material Table 1), giving confidence to our identified equation.

The credible intervals of all parameters cover the true value with the exception of $u_{xx}$ in the cases with 5\% noise and 2\% noise with 5\% missing data.
In addition, no extraneous terms are identified in any scenario.
The probability of including another term ($u_{xxx}$ for all four cases) is low, giving us relative certainty that the identified equation is indeed correct.
Clearly, the methodology eventually will fail when measurement error is too large or there is too much missing data. 
For example, we found when the measurement noise is greater than 8\% or more than 10\% of data are missing, we no longer recover the true model.

\begin{table}[t]
    \centering
    \begin{tabular}{c|c|c||l}
        Noise & Missing Data & Statistic & Discovered Equation \\
        \hline
        \hline
            &     & Mean      & $u_t = - 0.994 u u_x + 0.098 u_{xx}$\\
        0\% & 0\% & Lower HPD & $u_t = - 1.022 u u_x + 0.092 u_{xx}$\\
            &     & Upper HPD & $u_t = - 0.964 u u_x + 0.103 u_{xx}$\\
        \hline
            &     & Mean      & $u_t = - 0.990 u u_x + 0.096 u_{xx}$ \\
        2\% & 0\% & Lower HPD & $u_t = - 1.033 u u_x + 0.086 u_{xx}$ \\
            &     & Upper HPD & $u_t = - 0.954 u u_x + 0.103 u_{xx}$ \\
        \hline
            &     & Mean      & $u_t = - 0.981 u u_x + 0.094 u_{xx}$ \\
        5\% & 0\% & Lower HPD & $u_t = - 1.022 u u_x + 0.087 u_{xx}$ \\
            &     & Upper HPD & $u_t = - 0.951 u u_x + 0.099 u_{xx}$ \\
        \hline
            &     & Mean      & $u_t = - 0.957 u u_x + 0.087 u_{xx}$ \\
        2\% & 5\% & Lower HPD & $u_t = - 1.003 u u_x + 0.078 u_{xx}$ \\
            &     & Upper HPD & $u_t = - 0.931 u u_x + 0.095 u_{xx}$ \\
    \end{tabular}
    \caption{Discovered Burgers' equation (mean) and lower and upper HPD intervals with varying amounts of noise and missingness. The true Burgers' equation is $u_t = -uu_{x} + u_{xx}$.}
    \label{tab:burgers}
\end{table}

\subsection{Heat Equation}

\begin{figure}[t]
    \centering
    \includegraphics[width = 0.8\linewidth]{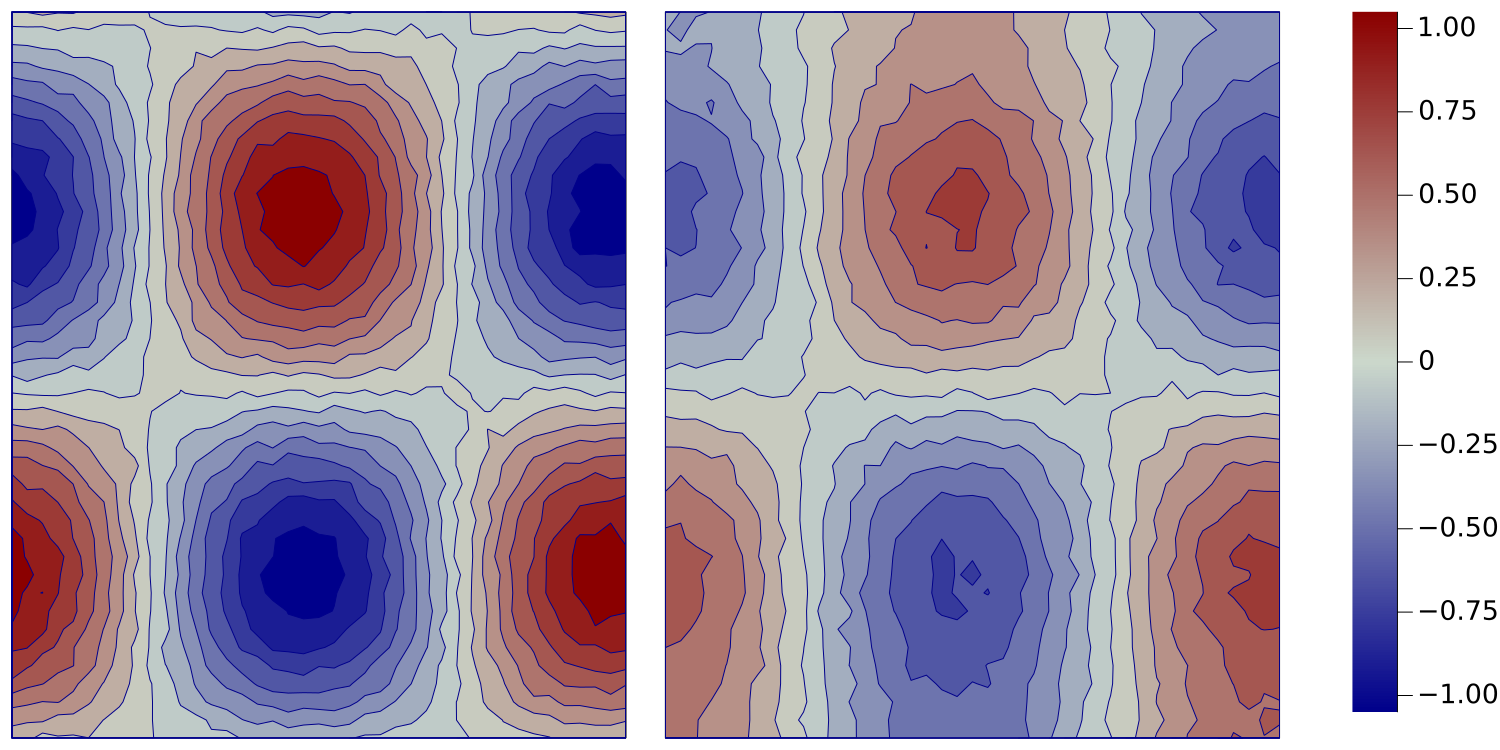}
    \caption{Data generated from the heat equation with 5\% noise at time step 0 (left, corresponding to $t = 0$) and time step 21 (right, corresponding to $t = 0.2$).}
    \label{fig:heat_data}
\end{figure}

The two-dimensional (2D) heat equation can be used to model the dissipation of heat over time.
We consider the 2D heat equation described by the PDE
\begin{align*}
    u_t(\vs,t) = \alpha \nabla^2 u(\vs,t) = \alpha u_{xx}(\vs,t) + \alpha u_{yy}(\vs,t),
\end{align*}
where $u(\vs,t)$ is the temperature of the surface at location $\vs = (x,y)$ and time $t$, $\alpha$ is the thermal diffusivity, and $\nabla^2 = u_{xx} + u_{yy}$ denotes the Laplacian operator.
Data are simulated using a central finite difference scheme over the spatial domain $D_s = [0,20] \times [0,20]$ with a spatial resolution of 0.5 for both the $x$ and $y$ directions, and over the time domain $D_t = [0, 2]$ with a temporal resolution of 0.01.
The the thermal diffusivity, $\alpha$, is set to 1.
The surface is initialized as
\begin{align*}
    u(\vs,0) = sin(2\pi x/40)*cos(2\pi y/40).
\end{align*}
We consider three cases -- no measurement noise, 2\% measurement noise, and 5\% measurement noise (Figure \ref{fig:heat_data}).

For all cases we specify our model with $P = 100 , Q = 80, |\mathcal{Z}| = 100, \kappa = 10^{-4}$ and define the feature library as
\begin{align*}
    [& u, u^2, u^3, u_x, u u_x, u^2 u_x, u^3 u_x, u_{xx}, u u_{xx}, u^2 u_{xx}, u^3 u_{xx}, u_y, u u_y, \\
    & u^2 u_y, u^3 u_y, u_{xy}, u u_{xy}, u^2 u_{xy}, u^3 u_{xy}, u_{yy}, u u_{yy}, u^2 u_{yy}, u^3 u_{yy}].
\end{align*}
Again, keeping terms with greater than 50\% inclusion probability, the recovered equation and 95\% HPD interval are shown in Table \ref{tab:heat}.
For all scenarios, we are able to correctly identify the true terms.
All HPD intervals except $u_{xx}$ for the scenario with 5\% noise and the scenario with 2\% noise and 5\% missing data cover the truth, and no extraneous terms are identified (Supplementary material Table 2).

\begin{table}[t]
    \centering
    \begin{tabular}{c|c||l}
        Noise & Statistic & Recovered Equation \\
        \hline
        \hline
            & Mean      & $u_t = 1.001 u_{xx} + 1.000 u_{yy}$ \\
        0\% & Lower HPD & $u_t = 0.999 u_{xx} + 0.998 u_{yy}$ \\
            & Upper HPD & $u_t = 1.002 u_{xx} + 1.002 u_{yy}$ \\
        \hline
            & Mean      & $u_t = 0.997 u_{xx} + 1.002 u_{yy}$ \\
        2\% & Lower HPD & $u_t = 0.991 u_{xx} + 0.995 u_{yy}$ \\
            & Upper HPD & $u_t = 1.002 u_{xx} + 1.009 u_{yy}$ \\
        \hline
            & Mean      & $u_t = 0.986 u_{xx} + 0.994 u_{yy}$ \\
        5\% & Lower HPD & $u_t = 0.967 u_{xx} + 0.977 u_{yy}$ \\
            & Upper HPD & $u_t = 1.000 u_{xx} + 1.016 u_{yy}$ \\
    \end{tabular}
    \caption{Discovered heat equation (mean) and lower and upper HPD intervals with varying amounts of noise where the true Heat equation is $u_t = u_{xx} + u_{yy}$.}
    \label{tab:heat}
\end{table}

\subsection{Reaction-Diffusion Equation}

The reaction-diffusion equation can be used to model the change in concentration or density of substances over time.
We consider the 2D reaction-diffusion parameterized by the PDE
\begin{align*}
    \vu_t(\vs, t) = \vD \nabla^2 \vu(\vs, t) + \vc(\vu(\vs,t)),
\end{align*}
where $\vu(\vs, t) = [u(\vs,t), v(\vs,t)]'$ may represent the concentration or density of two processes, $\vD$ is a diagonal matrix of the diffusion coefficient for each process, and $\vc(\cdot)$ is the (non)linear reaction function.
The reaction-diffusion equation can be used to model the interaction between a predator and prey population \citep{Hastings1996, Liu2019}.
To represent the interaction between prey and predator populations, we let $\vu = [u,v]'$ where $u$ and $v$ are the densities of the prey and predator populations, respectively.
We define $\vc(\cdot)$ to be the classic Lotka-Volterra model with a carrying capacity for the prey.
Specifically,
\begin{align*}
    \vc(\cdot) = 
    \begin{bmatrix}
        c_u(u(\vs, t), v(\vs, t)) \\
        c_v(u(\vs, t), v(\vs, t))
    \end{bmatrix} = 
    \begin{bmatrix}
        \gamma_0 u(\vs, t) - \frac{\gamma_0}{\gamma_1} u^2(\vs, t) - \beta u(\vs, t)v(\vs, t) \\
        \mu u(\vs, t) v(\vs, t) - \eta v(\vs, t)
    \end{bmatrix},
\end{align*}
where $\gamma_0$ is the prey growth rate, $\gamma_1$ is the prey carrying capacity, $\beta$ predation rate, $\mu$ is the predator growth rate, and $\eta$ is the predator death rate.

\begin{figure}[t]
    \centering
    \includegraphics[width = 0.8\linewidth]{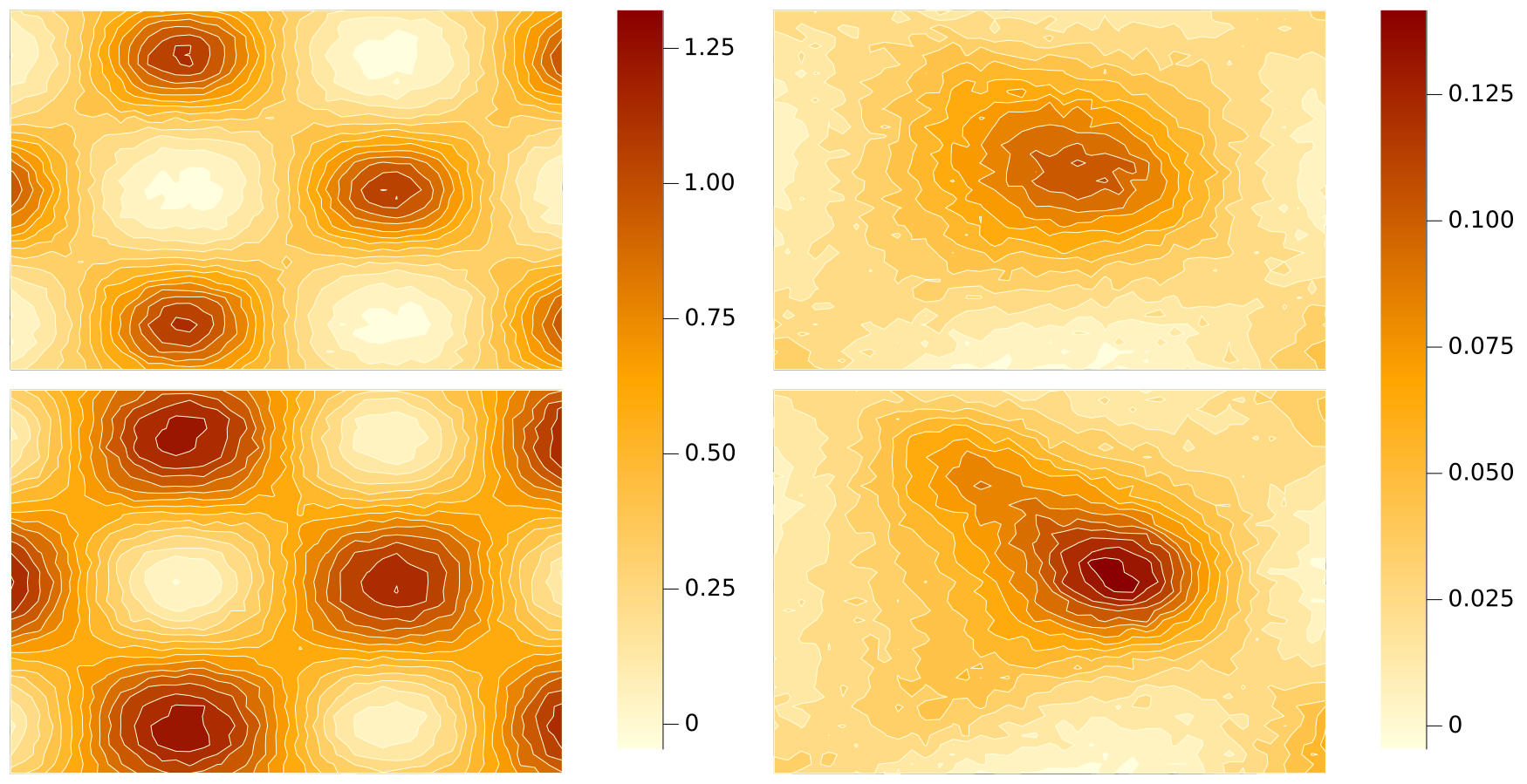}
    \caption{Data generated from the prey (left) and predator (right) reaction-diffusion system with 5\% measurement noise. Data are shown at time steps 11 (top, corresponding to $t=1$) and 31 (bottom, corresponding to $t=3$).}
    \label{fig:reaction_diffusion}
\end{figure}

Suppressing the spatial and temporal indices, the predator-prey reaction-diffusion equation is
\begin{align}\label{eqn:reaction_diffusion}
\begin{split}
    u_t & = D_u u_{xx} + D_u u_{yy} + \gamma_0 u - \frac{\gamma_0}{\gamma_1} u^2 - \beta u v \\
    v_t & = D_v v_{xx} + D_v v_{yy} + \mu u v - \eta v.
\end{split}
\end{align}
We simulate from (\ref{eqn:reaction_diffusion}) with $\gamma_0 = 0.4, \gamma_1 = 1.5, \beta = 0.5, \mu = 0.3, \eta = 0.1$ using a central finite difference scheme over the spatial domain $D_s = [-10, 10] \times [-10, 10]$ and the temporal domain $D_t = [0, 10]$ with a spatial and temporal resolution of $(0.5, 0.5)$ and 0.1, respectively.
The prey and predator densities are initialized as
\begin{align*}
    u(\vs,0) & = \exp\{cos(2 \pi x / 15)  sin(2 \pi y / 15)\} \\
    v(\vs,0) & = 0.1\exp\{cos(2 \pi y / 30 ) sin(2 \pi x / 30 - 5)\},
\end{align*}
respectively.
We again consider three scenarios -- no measurement noise, 2\% measurement noise, and 5\% measurement noise (Figure \ref{fig:reaction_diffusion}).

\begin{table}[t]
    \centering
    \small
    \begin{tabular}{c|c|c||l}
        Noise & Component & Statistic & Recovered Equation \\
        \hline
        \hline
            &      & Mean  & $u_t = 0.400 u - 0.266 u^2 - 0.500 u v + 0.099 u_{xx} + 0.100 u_{yy}$ \\
        0\% & Prey & Lower & $u_t = 0.399 u - 0.267 u^2 - 0.503 u v + 0.098 u_{xx} + 0.099 u_{yy}$ \\
            &      & Upper & $u_t = 0.400 u - 0.266 u^2 - 0.497 u v + 0.100 u_{xx} + 0.101 u_{yy}$ \\
        \hline
            &          & Mean  & $v_t = -0.100 v + 0.300 u v +  0.099 v_{xx} +  0.099 v_{yy}$ \\
        0\% & Predator & Lower & $v_t = -0.100 v + 0.299 u v +  0.099 v_{xx} +  0.099 v_{yy}$ \\
            &          & Upper & $v_t = -0.100 v + 0.300 u v +  0.100 v_{xx} +  0.100 v_{yy}$ \\
    \end{tabular}
    \caption{Discovered predator-prey reaction-diffusion equation (mean) and lower and upper HPD intervals with no noise. The true equations are $u_t = 0.4 u - 0.26 u^2 - 0.5 u v + 0.1 u_{xx} + 0.1 u_{yy}$ and $v_t = 0.3 u v - 0.1 v + 0.1 v_{xx} + 0.1 v_{yy}$.}
    \label{tab:rd_clean}
\end{table}

\begin{table}[t]
    \centering
    \small
    \begin{tabular}{c|c|c||l}
        Noise & Component & Statistic & Recovered Equation \\
        \hline
        \hline
            &      & Mean  & $u_t = 0.403 u - 0.270 u^2 - 0.492 u v + 0.144 u_{xx} + 0.145 u_{yy}$ \\
        2\% & Prey & Lower & $u_t = 0.400 u - 0.274 u^2 - 0.501 u v + 0.095 u_{xx} + 0.097 u_{yy}$ \\
            &      & Upper & $u_t = 0.405 u - 0.267 u^2 - 0.487 u v + 0.163 u_{xx} + 0.164 u_{yy}$ \\
        \hline
            &          & Mean  & $v_t = -0.098 v + 0.297 u v + 0.136 v_{xx} + 0.138 v_{yy}$ \\
        2\% & Predator & Lower & $v_t = -0.100 v + 0.296 u v + 0.095 v_{xx} + 0.099 v_{yy}$ \\
            &          & Upper & $v_t = -0.098 v + 0.300 u v + 0.161 v_{xx} + 0.155 v_{yy}$ \\
    \end{tabular}
    \caption{Discovered predator-prey reaction-diffusion equation (mean) and lower and upper HPD intervals with 2\% noise. The true equations are $u_t = 0.4 u - 0.26 u^2 - 0.5 u v + 0.1 u_{xx} + 0.1 u_{yy}$ and $v_t = 0.3 u v - 0.1 v + 0.1 v_{xx} + 0.1 v_{yy}$.}
    \label{tab:rd_noise}
\end{table}

\begin{table}[t]
    \centering
    \small
    \begin{tabular}{c|c|c||l}
        Noise & Component & Statistic & Recovered Equation \\
        \hline
        \hline
            &      & Mean  & $u_t = 0.401  u - 0.269 u^2 - 0.493 u v + 0.090 u_{xx} + 0.099 u_{yy}$ \\
        5\% & Prey & Lower & $u_t = 0.400  u - 0.270 u^2 - 0.504 u v + 0.085 u_{xx} + 0.091 u_{yy}$ \\
            &      & Upper & $u_t = 0.403  u - 0.267 u^2 - 0.487 u v + 0.093 u_{xx} + 0.102 u_{yy}$ \\
        \hline
            &          & Mean  & $v_t = -0.101 v + 0.300 u v + 0.092 v_{xx} + 0.106 v_{yy}$ \\
        5\% & Predator & Lower & $v_t = -0.103 v + 0.297 u v + 0.086 v_{xx} + 0.097 v_{yy}$ \\
            &          & Upper & $v_t = -0.099 v + 0.302 u v + 0.095 v_{xx} + 0.151 v_{yy}$ \\
    \end{tabular}
    \caption{Discovered predator-prey reaction-diffusion equation (mean) and lower and upper HPD intervals with 5\% noise. The true equations are $u_t = 0.4 u - 0.26 u^2 - 0.5 u v + 0.1 u_{xx} + 0.1 u_{yy}$ and $v_t = 0.3 u v - 0.1 v + 0.1 v_{xx} + 0.1 v_{yy}$.}
    \label{tab:rd_noise5}
\end{table}

For all cases we specify our model with $P = 225, Q = 40, |\mathcal{Z}| = 100, \kappa = [10^{-4}, 10^{-6}]$ and define the feature library as
\begin{align*}
    [u, u^2, u^3, v, v^2, v^3, uv, u^2v, uv^2, uu_x, uu_y, vv_x, vv_y, u_x, u_y, u_{xx}, u_{yy}, u_{xy}, v_x, v_y, v_{xx}, v_{yy}, v_{xy}].
\end{align*}
Posterior estimates of terms with greater than 50\% inclusion probability are shown in Tables \ref{tab:rd_clean}, \ref{tab:rd_noise}, and \ref{tab:rd_noise5} for the case with no noise, 2\% measurement noise, and 5\% measurement noise, respectively.
With no measurement noise, we see all terms are correctly identified and the 95\% HPD intervals all cover the truth.
For the scenario with 2\% noise, all terms are correctly identified and all coefficients except for $u^2$ in the prey equation cover the truth.
The scenario with 5\% measurement noise correctly identifies all terms and only $u^2$ and $u_{xx}$ for the prey equation and $v_{xx}$ for the predator equation have 95\% credible intervals that do not cover the truth.

\section{Barotropic Vorticity Equation}\label{sec:barotrophic}

Here we show the ability of our model to discover governing dynamic equations using real world data.
The 500-hPa level of the atmosphere is often known as the ``level of non-divergence'' because in the absence of strong cyclogenesis, the flow is essentially horizontal and non-divergent at this mid-level of the atmosphere.
Such flows can often be modeled quite effectively with ``barotropic'' dynamics (in a barotropic fluid, the density is constant along a constant pressure surface).
Indeed, the first successful numerical weather forecasts were based on the advection of relative vorticity (rotation of the fluid in the horizontal dimension) at the 500-hPa level using the so-called barotropic vorticity equation \citep[BVE;][]{Charney1950}.
The BVE is given as
\begin{align*}
    \xi_t(\vs, t) = -\vv(\vs, t) \cdot \nabla(\xi(\vs, t) + f(\phi(\vs))),
\end{align*}
where $\xi = \frac{\partial v}{\partial x} - \frac{\partial u}{\partial y}$ is the relative vorticity, $\vv = (u, v)$ is the non-divergent horizontal wind vector with $u$ the wind component in the zonal (east-west) direction, $v$ the meridional (north-south) wind component, and $f = 2\Omega sin(\phi)$ is the Coriolis parameter with $\Omega = 7.292 \times 10^{-5}$ rad $s^{-1}$ the angular speed of rotation of the Earth (note -- not to be confused with the dynamic discovery library as defined previously), and $\phi(\vs)$ the latitude in radians at location $\vs$.
The relative vorticity and non-divergent wind components can each be written in terms of the stream function, $\psi$.
In particular, the vorticity is given in terms of the Laplacian of the streamfunction, $\xi =  \frac{\partial^2 \psi}{\partial x^2} + \frac{\partial^2 \psi}{\partial y^2}$, and the wind components are given by $v = \frac{\partial \psi}{\partial x},$ and $u = -\frac{\partial \psi}{\partial y}$  \citep[e.g.,][]{Holton2012}.
We can substitute these into the BVE to get an alternate form (suppressing the spatial and temporal indices):
\begin{align}\label{eqn:streamfunction}
    \nabla^2\psi_{t} = \psi_y\psi_{xxx} + \psi_y\psi_{xyy} - \psi_x\psi_{xxy}-\psi_x\psi_{yyy}-\psi_xf_y,
\end{align}
where $\nabla^2\psi_{t} = \psi_{xxt} + \psi_{yyt}$. This form is very useful for dynamic discovery because we need only consider functions of $\psi$. The first four terms on the RHS correspond to the advection of relative vorticity and the last term on the RHS corresponds to the advection of planetary vorticity (that associated with planetary rotation).
Furthermore, note that the streamfunction can be computed based on the observed geopotential height, $\Phi$, where $\psi(\vs, t) = \Phi^*(\vs, t) / f(\phi(\vs))$, $\Phi^*(\vs, t) = \Phi(\vs, t) - \int_D \Phi(\vs, t)$, and $D$ is the observed domain.

\begin{figure}[ht]
    \centering
    \includegraphics[width = 0.8\linewidth]{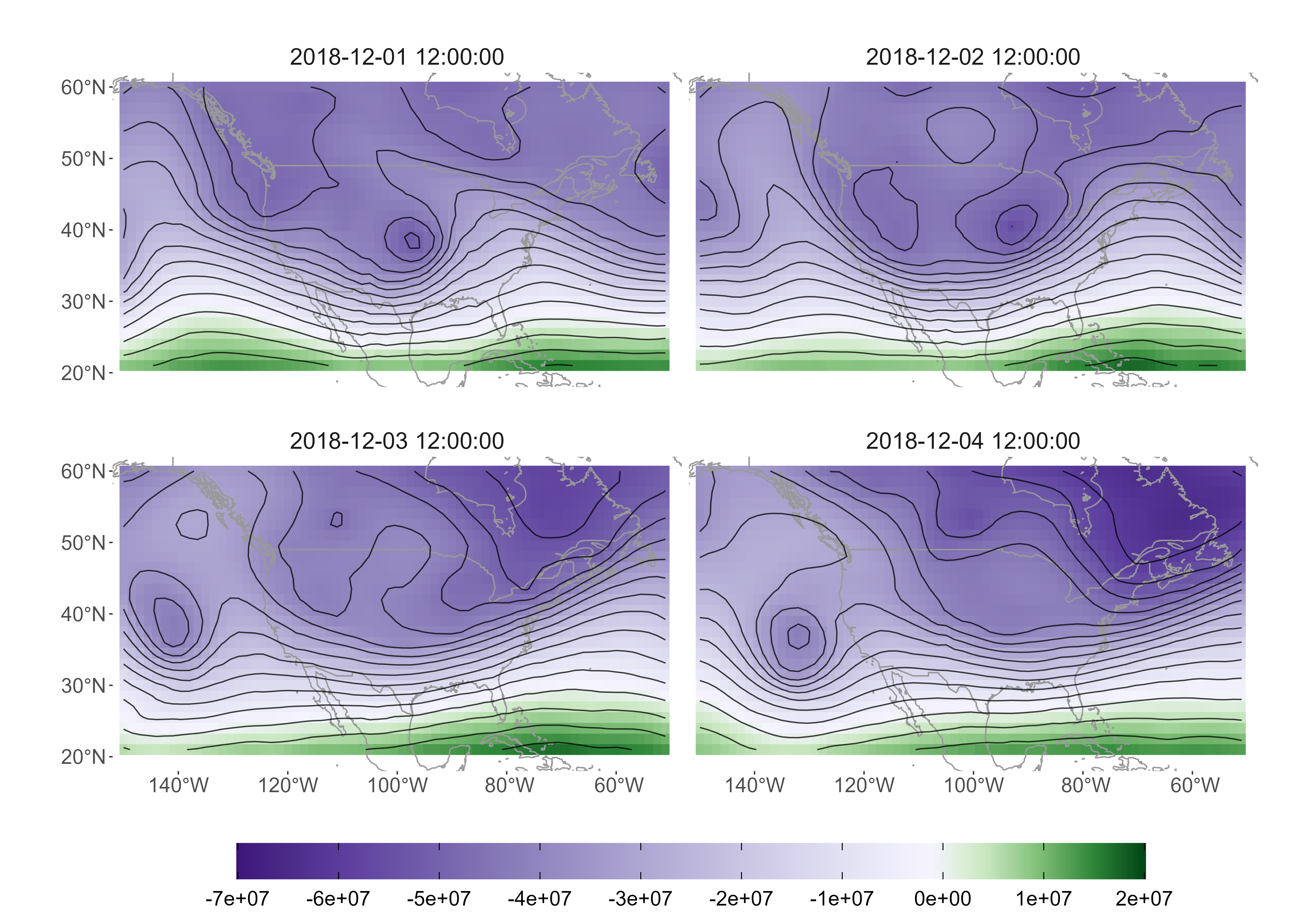}
    \caption{Streamfunction ($m^2/s$) data at 12:00 UTC on November 1, 2, 3, and 4, 2018. Red and blue correspond to lower value and upper values, respectively, with contour lines included for visual aid.}
    \label{fig:stream}
\end{figure}

Here, we use hourly data generated using Copernicus Atmosphere Monitoring Service information (2022)\footnote{\url{https://cds.climate.copernicus.eu/cdsapp\#!/dataset/10.24381/cds.bd0915c6?tab=overview}} of relative geopotential height at 500-hPA.
Data are collected hourly from December 1st to December 31, 2018 over the spatial domain $D_s = [-150^{\circ}W, -50^{\circ}W] \times [20^{\circ}N, 60^{\circ}N]$ at a resolution of 1.5 degrees (see Figure \ref{fig:stream} for example plots), resulting in $(67 \times 27) \times 744$ space-time locations.
We compute the streamfunction from the geopotential height and use this as the observed data for the discovery.
We also compute the derivative of the Coriolis parameter in the latitudinal direction, $f_y = 2\Omega cos(\phi(\vs))$, and use it as a covariate in the model (i.e., $\omega(\vs, t) = f_y(\vs)$ from (\ref{eqn:gen_process_model})).

We specify our model with $P = 200, Q = 300, |\mathcal{Z}| = 200, \kappa = 10^{-4}$ and define the feature library as
\begin{align*}
    [\psi, \psi_x, \psi_{xx}, \psi_y, \psi_{yy}, \psi_{xy}, & \psi_x \psi_{xxx}, \psi_y \psi_{xxx}, \psi_x \psi_{yyy}, \psi_y \psi_{yyy}, \\
    \psi_x \psi_{xxy}, \psi_y \psi_{xxy}, \psi_x & \psi_{xyy}, \psi_y \psi_{xyy}, \psi_x f_y, \psi_y f_y].
\end{align*}
Posterior estimates for the recovered equation are shown in Table \ref{tab:streamfunction}, where only terms with a posterior inclusion probability greater than 50\% were kept.
While we do not know the true equation in this case (because the barotropic vorticity equation is only an approximation of the dynamics in the atmosphere), we see the discovered equation closely resembles the hypothesized BVE given in (\ref{eqn:streamfunction}).

Although the sign for each discovered term aligns with the sign in the BVE, we note that the coefficient values are different from the hypothesized equation and the planetary vorticity term  ($\psi_x f_y$) is not significant.
There are several reasons why this is likely the case and not surprising.
First, it is possible that this particular time period is not barotropic (e.g., baroclinic), which would require library terms our current framework cannot accommodate (e.g., flow in the vertical direction, density, temperature).
Such transient changes from barotropic to baroclinic flow regimes are common in the upper-level atmospheric flow.
The lack of significance of the planetary vorticity term is also interesting.
Recall that the meridional velocity is related to the gradient of streamfunction in the meridional direction, so this term can be alternatively written $\psi_x f_y = v f_y$.
Thus, if the flow is primarily zonal (east-west), the north-south flow component ($v$) is quite small and this term may not be important for a particular time period.
Indeed, visual examination of the data suggests that the flow is dominated by zonal flow and this is further confirmed seeing that the magnitude of the planetary vorticity term in the BVE is an order of magnitude smaller compared to the relative vorticity terms.
Yet, the planetary vorticity term is still important to the advection of relative vorticity because of the Coriolis effect on the flow, resulting in the term being identified as important (if not significant).
We also note that the planetary vorticity term in the BVE is important for supporting so-called Rossby waves (slowly moving westward propagating synoptic-scale waves) in the upper level flow. 
Yet, the time period and spatial extent of the data considered here are likely not sufficient to capture the full spatial and temporal scale of Rossby waves  \citep[e.g.,][]{Holton2012}.
Thus, it is somewhat remarkable that the dynamic discovery methodology was still able to infer realistic properties of the system of interest that generally align with the theoretical system. 
To the best of our knowledge, this is the first time that data-driven discovery methods have been applied to real-world atmospheric data and have identified physically plausible features.

\begin{table}[ht]
    \small
    \centering
    \begin{tabular}{c||l}
        Statistic & Discovered Equation \\
        \hline
        \hline
        Mean      & $\nabla^2 \psi_t = 0.289 \psi_{y} \psi_{xxx} + 0.277 \psi_{y} \psi_{xyy} - 0.280 \psi_{x} \psi_{xxy} - 0.185 \psi_{x} \psi_{yyy} - 6.354 \psi_x f_y$ \\
        Lower HPD & $\nabla^2 \psi_t = 0.235 \psi_{y} \psi_{xxx} + 0.267 \psi_{y} \psi_{xyy} - 0.343 \psi_{x} \psi_{xxy} - 0.215 \psi_{x} \psi_{yyy} - 7.570 \psi_x f_y$ \\
        Upper HPD & $\nabla^2 \psi_t = 0.317 \psi_{y} \psi_{xxx} + 0.286 \psi_{y} \psi_{xyy} - 0.223 \psi_{x} \psi_{xxy} - 0.160 \psi_{x} \psi_{yyy} + 1.491 \psi_x f_y$ \\
    \end{tabular}
    \caption{Discovered equation for the BVE (mean) and lower and upper HPD intervals where the theoretical BVE is $\nabla^2\psi_{t} = \psi_y\psi_{xxx} + \psi_y\psi_{xyy} - \psi_x\psi_{xxy}-\psi_x\psi_{yyy}-\psi_xf_y$.}
    \label{tab:streamfunction}
\end{table}

\section{Conclusion}\label{sec:conclusion}

We have proposed a data-driven approach for learning complex non-linear spatio-temporal dynamic equations that is robust to measurement noise and missing data.
Our approach uses a Bayesian hierarchical model where the dynamic equation is embedded in the latent process enabling the discovery of dynamic equations within the statistical framework.
Additionally, the model provides probabilistic estimates of inclusion for each component of the feature library and estimates of uncertainty for the recovered parameters, giving a deeper understanding to the dynamic system.
This all stems from the expansion of the dynamic process in terms of basis functions, bypassing the need for numerical differentiation and enabling the estimate of the derivatives within a probabilistic framework.

Improvements to the current framework should focus on the specification of the feature library and the selection prior.
A feature library that is uninhibited by the users choice (i.e., the model could generate library terms) would remove user bias.
This is akin to what has been proposed with symbolic regression.
Also, a different choice in selection prior for the coefficients, perhaps one that also penalizes model complexity, has the potential to improve selection performance.
While there are a variety of selection priors in the literature, a prior directed towards this problem will provide noticeable improvement on the identification of the system.

From the application perspective, the promising and intriguing results from the BVE example suggest further investigation.
In particular, it would be informative to consider larger spatial domains (hemispheric) and temporal time periods (many months) to determine if the presence of Rossby waves are sufficient to lead to a significant planetary vorticity term.
This also suggests experiments with numerical simulation models that can easily be controlled to exhibit barotropic or baroclinic flow.
Such models will allow us to investigate the effects of the flow regime on the ability to detect realistic dynamics.

\setstretch{1}
\bibliographystyle{apalike}
\bibliography{references}

\setcounter{section}{0}
\renewcommand{\thesection}{S}

\section{Supplementary Material}

\subsection{Inclusion Probabilities for Spike and Slab}\label{sec:sas}

Let the residual sum of squares for the model with the $M(n,d)$ term included be $RSS_{\gamma}$ and the model without the $M(n,d)$ term included be $RSS_{\setminus\gamma}$.
The probability any element is included is given as
\begin{align*}
    p(\vgamma(n,d)=1|\cdot) = \frac{1}{1 + \frac{1-\pi(n)}{\pi(n)}R_{\gamma}(n,d)}
\end{align*}
where
\begin{align*}
    R_{\gamma}(n,d) = (g+1)^{1/2} \left(\frac{RSS_{\gamma}}{RSS_{\setminus\gamma}}\right)^{ST/2-1}.
\end{align*}
Denote $\beta = \frac{RSS_{\gamma}}{RSS_{\setminus\gamma}}$ and solving for the number of observations $ST$, 
\begin{align*}
    n_{obs} \coloneqq ST = 2\left( \log \left( \frac{R_{\gamma}(n,d)}{(g+1)^{1/2}} \right) \Big/ \log(\beta) + 1 \right).
\end{align*}

We then use the value of $n_{obs}$ to inform our subsample size based on $g$, the ratio of the RSSs, and an informed value of $R$.
For example, assume $\pi = 0.5$ such that every parameter has a 50\% chance of being included in the model.
We would take $R = 1$ resulting in $p(\vgamma(n,d)=1|\cdot) = 0.5$.
$n_{obs}$ is then chosen by solving the equation under a hypothetical $\beta$ (e.g., $0.99$ or $0.95$).

In choosing the value $\beta$ there are a couple things to consider.
If the terms in the library are highly correlated, there is likely to be confounding and incorrect variables may have a larger impact on the RSS.
This issue is detected using the condition number of the correlation matrix (e.g., $\vF^{*'} \vF^*$ where $\vF^*$ is the normalized version of $\vF$), where a large condition number (e.g., greater than 1000) indicates multicollinearity.
Under the scenario where the condition number is large, $\beta$ should be chosen to be smaller (i.e., 0.9 or 0.95), resulting in a smaller subsampled size.
Alternatively, if the variables are less correlated (i.e., condition number smaller than 1000), the impact of an incorrect variable on the RSS will be less.
In this case $\beta$ can be chosen to be closer to one (e.g., 0.99 or 0.999), resulting in a larger subsample size.
When fitting the model, the parameter $\pi$ will be estimated and will (likely) not be 0.5.
However, empirically we have found taking $\pi = 0.5$ and $R = 1$ to solve for $obs$ works well.

For the simulations, the condition number of the correlation matrix for Burgers' is approximately 28940, the Heat equation is approximately 1840, and the reaction-diffusion equation is approximately 6510.
We chose $\beta$ to be 0.9, 0.99, and 0.95 for Burgers', the Heat, and the reaction-diffusion equations, respectively.
For the real-world example, the condition number of the correlation matrix is approximately 480 and we chose $\beta$ to be 0.99.

A parallel can be drawn between the subsampling and sequential thresholded least squares \citep[STLS;][]{Brunton2016} or sequential threshold ridge regression \citep[STRidge;][]{Rudy2017}.
Because the inclusion probability is affected by the subsample size (i.e., reduces the inclusion probability or highly unlikely terms), this is analogous to a probabilistic extension of the thresholding approaches.
That is, instead of assigning a hard threshold, where values less than a predetermined value are set to zero, the subsampling approach impacts the probability of a variable being included below a certain threshold based on a specified $\beta$.

\subsection{Sampling Algorithm}\label{sec:algorithm}

To simplify notation, denote $\vB_0(\vs, t) = \vphi_{t^{(0)}}(t) \otimes \vpsi(\vs)$ and $\vB_J(\vs, t) = \vphi_{t^{(J)}}(t) \otimes g(\vpsi(\vs))$.
At time $t$ and location $\vs$, the full model is

\begin{align*}
    \vv(\vs, t) & \sim N(\vTheta \vA \vB'_0(\vs, t), \vSigma_{V}(\vs, t)) \\
    \vTheta \vA \vB'_J(\vs, t) & \sim N(\vM \vf(\vs, t), \vSigma_{U}) \\
    \vM_{nd}|\vgamma_{nd} & = (1-\vgamma_{nd})\delta_0 + \vgamma_{nd} N(0, c) \\
    p(\vgamma_{nd} = 1|\pi_n) & = \pi_n \\
    \pi & \sim Beta(a, b) \\
    \vSigma_{V} & = \vH(\vs, t) diag(\sigma_{V1}^2, ..., \sigma_{VN}^2) \vH'(\vs, t) \\
    \sigma_{Vn}^2 & \sim IG(\nu_V/2, \nu_y/a_{Vn}) \\
    \vSigma_{U} & = diag(\sigma_{V1}^2, ..., \sigma_{VN}^2) \\
    \sigma_{Un}^2 & \propto 1/\sigma_{Un}^2.
\end{align*}

For iteration $\ell = 1, ..., \mathcal{L}$ do:

\begin{enumerate}
    \item Obtain minibatch $\mathcal{Q}$. 
    
    \item Update $\vgamma$:
        Denote $\vf_{\vgamma}$ as the design matrix consisting only of columns of $\vf$ corresponding to non-zero effects, $\vG_{\vgamma} = \frac{g}{g+1}(\vf_{\vgamma}'\vf_{\vgamma})^{-1}, \vg_{\vgamma} = \vG_{\vgamma}\vf_{\vgamma}'\vTheta \vA(n) \vB'_J$, and $\vy_{\vgamma} = \frac{1}{2}((\vTheta \vA(n) \vB'_J)'(\vTheta \vA(n) \vB'_J) - \vg_{\vgamma}'\vG_{\vgamma}^{-1}\vg_{\vgamma})$.
        Let $\vG_{\vgamma,0}, \vg_{\vgamma,0}$, and $\vy_{\vgamma,0}$ correspond to $\vgamma(d) = 0$ and $\vG_{\vgamma,1}, \vg_{\vgamma,1}$, and $\vy_{\vgamma,1}$ correspond to $\vgamma(d) = 1$.
        Sample each element of $\gamma_{nd}, n = 1, ..., N, d = 1, ..., D$ of the indicator vector $\vgamma$ from
        \begin{align*}
            p(\gamma_{nd} = 1|\vgamma_{\setminus nd}, \vTheta, \vA(n), \vB'_J) = \frac{1}{1 + R_{\gamma}(n,d)\frac{1-\pi_n}{\pi_n}},
        \end{align*}
        where $R_{\gamma}(n,d) = (g+1)^{1/2} \frac{\vy_{\vgamma,1}}{\vy_{\vgamma,0}}^{ST/2 - 1}$.

    \item Update $\vpi$: For $n = 1, ..., N$, sample
    \begin{align*}
        [\pi_n|\cdot] \sim Beta\left(a + \sum_d \vgamma_{nd}, b + D - \sum_d \vgamma_{nd}\right).
    \end{align*}
    
    \item Update $\vM$: For $n = 1, ..., N$, set $\vM_{nd} = 0$ is $\vgamma_{nd} = 0$. For non-zero elements, sample 
    \begin{align*}
        [\vM_n|\cdot] \sim Gau\left(\vg_{\vgamma}, \sigma^2_{Un}\vG_{\vgamma}\right).
    \end{align*}
    
    \item Update $\vSigma_U$: For $n = 1, ..., N$, sample
    \begin{align*}
        [\sigma^2_{Un}|\cdot] \sim IG\left(\frac{N-1}{2}, \frac{1}{2}\left((\vTheta \vA_{(3)} \vB'_J(\vs, t))'(\vTheta \vA_{(3)} \vB'_J(\vs, t)) - \vg_{\vgamma}'\vG_{\vgamma}^{-1}\vg_{\vgamma}\right)\right).
    \end{align*}
    
    \item Update $\vSigma_V$: For $n = 1, ..., N$, sample
    \begin{align*}
        [\sigma^2_{Vn}|\cdot] \sim IG\left( \frac{TS + \nu_V}{2}, \frac{\nu_V}{a_V} + \frac{1}{2} \sum_{t=1}^T \sum_{s=1}^S (V(\vs, t, n) - \vtheta_n \vA B'_0(\vs, t))(V(\vs, t, n) - \vtheta_n \vA B'_0(\vs, t))' \right)
    \end{align*}
    and
    \begin{align*}
        [a_{Vn}|\cdot] \sim IG\left( \frac{\nu_V + 1}{2}, \frac{\nu_V}{\sigma^2_{Vn} + \frac{1}{A^2_V}} \right)
    \end{align*}
    
    \item Update $\vA$: Use (4.1) to update $\vA$.
    
\end{enumerate}

\subsection{Proofs of Propositions}\label{sec:props}

\textbf{Proposition 1.} \textit{The mode-3 decomposition of $[\![ \EuScript{A}; \vPsi, \vPhi_{t^{(J)}}, \vTheta ]\!] = \EuScript{F} \times_3 \vM + \widetilde{\veta}$ where $\veta(\vs,t) \overset{i.i.d.}{\sim} N_N(\vec{0}, \Sigma_U)$ in space and time at location $\vs$ and time $t$ is
\begin{align*}
    \vTheta \vA (\vphi_{t^{(J)}}(t) \otimes \vpsi(\vs))' = \vM \vf(\vA, \vpsi(\vs), \vpsi_x(\vs), \vpsi_y(\vs), \vpsi_{xy}(\vs), ..., \vphi_{t^{(0)}}(t), ..., \vphi_{t^{(J)}}(t), \vomega(\vs,t)) + \veta(\vs,t),
\end{align*}
where $\vA$ is a $R \times PQ$ matrix of basis coefficients, $\vpsi(\vs)$ is a length-$P$ vector of spatial basis functions, $\vphi(t)$ is a length-$Q$ vector of temporal basis functions, and $\vTheta$ is a $N \times R$ matrix of component basis functions.}

\begin{quote}
    \normalsize{\textbf{Proof of Proposition 1.} For the LHS, $\vTheta \vA (\vPhi_{t^{(J)}} \otimes \vPsi)'$ is the mode-3 matricization of $\EuScript{A} \times_1 \vPsi \times_2 \vPhi_{t^{(J)}} \times_3 \vTheta$ \citep[see][for a proof of this property]{Kolda2006}.
    For the RHS, from the property of the n-mode product, $\EuScript{F} \times_3 \vM = \vM \vf(\cdot)$, where $\vf(\cdot)$ is the mode-3 matricization of $\EuScript{F}$.
    The arguments of $\vf(\cdot)$, namely $\vU, \vU_x, \vU_y, ..., \vU_{t^{(0)}}, ..., \vU_{t^{(J-1)}}$, are represented using their basis expansion, resulting in $\vf(\cdot)$ depending on $\vPsi, \vPhi, \vTheta, \EuScript{A}$ and any derivatives of $\vPsi$ and $\vPhi$ needed for the library.
    The value at a specific space-time location is determined from the $\vs$th and $t$th column of $\vPsi$ and $\vPhi$, respectively.
    The last term on the RHS, $\veta(\vs,t)$, is a the mode-3 matricization of the uncertainty tensor $\widetilde{\veta}$, where each space-time location has the same variance-covariance matrix $\Sigma_U$.}
\end{quote}

\noindent \textbf{Proposition 2.} \textit{Let $g(\cdot)$ be a linear differential operator.
The basis formulation of a PDE with a space-time response $g(\vu_{t^{(J)}}(\vs, t))$ is 
\begin{align*}
    \vTheta \vA (\vphi_{t^{(J)}}(t) \otimes g(\vpsi(\vs)))'.
\end{align*}}

\begin{quote}
    \normalsize{\textbf{Proof of Proposition 2.} Let $g(\EuScript{U}) = \{g(u(\vs, t, n)): \vs \in D_s, t = 1, ..., T, n = 1, ..., N\}$ be a function of the tensor of the continuous process observed at discrete space-time locations.
    Decomposing $g(\EuScript{U})$ in terms of spatial, temporal, and component basis functions, $g(\EuScript{U}) \approx g([\![ \EuScript{A}; \vPsi, \vPhi_{t^{(J)}}, \vTheta ]\!])$.
    From Proposition 1, the mode-3 basis decomposition of $g([\![ \EuScript{A}; \vPsi, \vPhi_{t^{(J)}}, \vTheta ]\!])$ is $g(\vTheta \vA (\vPhi_{t^{(J)}} \otimes  \vPsi))'$.
    By linearity of $g(\cdot)$ and properties of the Kronecker product, 
    \begin{align*}
        g(\vTheta \vA (\vPhi_{t^{(J)}} \otimes  \vPsi))' = \vTheta \vA (\vPhi_{t^{(J)}} \otimes  g(\vPsi))'.
    \end{align*}
    The function at location $\vs$ and time $t$ is $\vTheta \vA (\vphi_{t^{(J)}}(t) \otimes  g(\vpsi(\vs)))'.$}
\end{quote}

\FloatBarrier

\subsection{Extra Tables}\label{sec:tabs}

\begin{table}[h]
    \centering
    \begin{tabular}{c|c||c|l||c|l}
        Noise & Missing Data & First Term & Probability & Second Term & Probability \\
        \hline
        \hline
        0\% & 0\% & $u_{xxx}$ & 0.193 & $uu_{xxx}$ & 0.139 \\
        \hline
        2\% & 0\% & $u_{xxx}$ & 0.222 & $uu_{xxx}$ & 0.145 \\
        \hline
        5\% & 0\% & $u_{xxx}$ & 0.235 & $uu_{xxx}$ & 0.164 \\
        \hline
        2\% & 5\% & $u_{xxx}$ & 0.229 & $uu_{xxx}$ & 0.150
    \end{tabular}
    \caption{Feature library term with the highest (first term) and next highest (second term) probability of inclusion that was not included in the discovered equation for data generated from Burgers' equation.}
    \label{tab:burgers_next}
\end{table}

\begin{table}[h]
    \centering
    \begin{tabular}{c||c|l||c|l}
        Noise & First Term & Probability & Second Term & Probability \\
        \hline
        \hline
        0\% & $u^3u_{xy}$ & 0.050 & $u^2u_{xy}$ & 0.037 \\
        \hline
        2\% & $u$ & 0.015 & $u^2u_{xx}$ & 0.013 \\
        \hline
        5\% & $u$ & 0.030 & $u^2u_{xx}$ & 0.010
    \end{tabular}
    \caption{Feature library term with the highest (first term) and next highest (second term) probability of inclusion that was not included in the discovered equation for data generated from the heat equation.}
    \label{tab:heat_next}
\end{table}

\begin{table}[h]
    \centering
    \begin{tabular}{c|c||c|l||c|l}
        Noise & Component & First Term & Probability & Second Term & Probability \\
        \hline
        \hline
        0\% & Prey & $u^3$ & 0.040 & $u_{xy}$ & 0.039\\
        \hline
        0\% & Predator & $u^2 v$ & 0.057 & $u_{xx}$ & 0.036 \\
        \hline
        2\% & Prey & $vv_y$ & 0.022 & $v_{yy}$ & 0.020 \\
        \hline
        2\% & Predator & $v^3$ & 0.020 & $vv_x$ & 0.019 \\
        \hline
        5\% & Prey & $v_{xx}$ & 0.029 & $v_{x}$ & 0.027 \\
        \hline
        5\% & Predator & $u^2$ & 0.024 & $u$ & 0.023
    \end{tabular}
    \caption{Feature library term with the highest (first term) and next highest (second term) probability of inclusion that was not included in the discovered equation for data generated from the reaction-diffusion equation.}
    \label{tab:rd_next}
\end{table}

\section*{Acknowledgments}
The authors would like to acknowledge Dr. Ralph Milliff for comments on an early draft and for helpful discussions concerning the results from the barotropic vorticity example.
This research was partially supported by the U.S.~National Science Foundation (NSF) grant SES-1853096 and the U.S. Geological Survey Midwest Climate Adaptation Science Center (CASC) grant No.G20AC00096. 



\end{document}